\documentstyle[preprint,aps,prb]{revtex}
\begin{document}
\draft
\tightenlines

\title{Thomas-Fermi-Dirac-von Weizs\"acker hydrodynamics in \\
laterally modulated electronic systems}

\author{B.P.~van Zyl and E.~Zaremba}
\address{Department of Physics, Queen's University,\\ Kingston, 
Ontario, Canada K7L 3N6}
\date{\today}

\maketitle

\begin{abstract}
We have studied the collective plasma
excitations of a two-dimensional electron gas with an {\em arbitrary}
lateral charge-density modulation. The dynamics is formulated
using a previously developed hydrodynamic theory based on the
Thomas-Fermi-Dirac-von Weizs\"acker approximation. In this
approach, both the equilibrium and dynamical properties
of the periodically modulated electron gas are treated in a
consistent fashion. We pay particular attention to the 
evolution of the collective excitations as the system undergoes
the transition from the ideal two-dimensional limit to the
highly-localized one-dimensional limit.
We also calculate the power absorption in the long-wavelength limit to
illustrate the effect of the modulation on the modes probed by 
far-infrared (FIR) transmission spectroscopy.
\end{abstract}

\section{INTRODUCTION}

Two-dimensional electronic systems, as found in single
heterojunction interfaces or in more complex structures such as
quantum wells, have been studied intensively because of their
fundamental interest and for their potential 
application\cite{review1,review2,review3,review4}.
In particular, the two-dimensional electron gas (2DEG) with a 
spatially periodic modulation of its equilibrium density has attracted 
considerable attention both theoretically\cite{krasheninnikov81,dassarma85,eliasson86a,eliasson86b,lai86,cataudella88,zhu88,que89,wulf90a,wulf90b,dahl90,shikin90,park92,schaich92,gudmundsson95}
and experimentally\cite{mackens84,hansen87,brinkop88,demel88,drexler92,hertel94,steinbach96,heitmann97,frank97}. 
One way of achieving these systems\cite{mackens84} 
is to apply a voltage between a patterned gate
and the 2DEG, thereby inducing a one-dimensional periodic
modulation of the density. As the gate voltage is increased, electrons
tend to `pool' near the minima of the modulating potential and, in the
limit of strong modulation, the system transforms into a periodic array
of 1D quantum wires. It is clear that the restricted motion of the
electrons along the wires will have a profound effect on the character
of the plasma excitations in the system. Our main focus in this paper 
will be to study the evolution of these collective modes with 
increasing modulation of the 2DEG, and in particular, the crossover 
from 2D to 1D behaviour. 

The theoretical approaches commonly used to discuss collective
excitations in charge-density-modulated systems can be classified
broadly as being either hydrodynamic in nature\cite{eliasson86a,eliasson86b,lai86,cataudella88,forstmann86,fetter74}, 
or based on the random phase approximation (RPA)\cite{dassarma85,zhu88,que89,wulf90a,wulf90b,dahl90,park92,schaich92,gudmundsson95}. 
The hydrodynamic approaches are appealing because
of their relative mathematical and computational simplicity. Although
they vary in their level of sophistication, they all have in common the
objective of describing the dynamics of the electronic system in terms
of a closed set of equations for the density and velocity fields. At the
simplest level of approximation\cite{krasheninnikov81,cataudella88,shikin90}, 
the current density in the electron
fluid is determined by a local conductivity, proportional to the
equilibrium electron density. Poisson's equation is then used to
relate the electric field driving the current to the fluctuating 
electron density. Rather than explicitly determining the equilibrium 
properties of the
electronic system, the equilibrium density is simply chosen
to have some physically reasonable form. We shall refer to such theories
as `classical' in that no explicit quantum mechanical aspects are
included in either the equilibrium or dynamical properties.

Although the classical treatment is useful for obtaining a qualitative
understanding of the collective modes in a modulated 2DEG, it has the
shortcoming of not being able to account for nonlocal effects (e.g. in
the magneto-conductivity tensor) which become important with increasing
magnitude of the plasmon wavevector. Nonlocal effects in the plasmon
dispersion can be included in the hydrodynamic theory through the
introduction of an electronic compressibility\cite{lai86,forstmann86}. 
In this way some information about the quantum 
mechanical equation of state can be built into the
theory. In this respect, these theories can be referred to as
`semiclassical'. The RPA on the other hand, provides a
fully quantum mechanical description of both the equilibrium and
dynamical behaviour. However, being more general, the
computational demands for its implementation are
significantly greater. Furthermore, since it
includes both single-particle and collective aspects, it is sometimes
difficult to cleanly identify excitations which are predominantly
collective in nature\cite{schaich94}. For both of these reasons, it 
is still worthwhile having available hydrodynamic based theories.

In this paper, we investigate the collective response of a modulated
2DEG using the Thomas-Fermi-Dirac-von Weizs\"acker (TFDW) hydrodynamics
previously developed to treat magnetoplasma excitations
in three-dimensional parabolic wells\cite{zaremba94,zaremba98}
and electron rings\cite{zaremba96}. One of the
primary virtues of this approach is that it is based on a reasonably
accurate description of the ground state properties of the electronic
system, which are determined self-consistently from the minimization of
the TFDW energy functional. Thus, unlike most earlier hydrodynamic 
treatments, the form of the equilibrium density is not chosen 
arbitrarily. Perturbations of the system away from the equilibrium 
state generate
internal forces which drive the system back towards equilibrium.
These forces are consistently included in the TFDW hydrodynamic 
equations used to describe the dynamics of the system. Although
somewhat more sophisticated, the present approach nevertheless
retains much of the mathematical simplicity of the usual hydrodynamic 
theories.

Our paper is organized as follows. In Sec.~\ref{equilibrium} we 
determine the equilibrium properties  of a periodically modulated 
2DEG within the TFDW approximation.  Sec.~\ref{collective} 
then provides the mathematical formalism needed to study the 
dynamics of the collective modes and includes a derivation of
the power absorption which is typically measured in infrared 
transmission experiments.  In Sec.~\ref{2dto1d}, 
we consider in detail the crossover from 2D to 1D behavior in the 
periodically-modulated 2DEG, and finally in Sec.~\ref{conclusions},
we present our concluding remarks.

\section{EQUILIBRIUM PROPERTIES}
\label{equilibrium}

The fabrication of mesoscopic devices often begins with a
2DEG in
metal-oxide-semiconductor (MOS) structures or epitaxially grown
heterojunctions.  In these systems, the strong confinement of
the electrons in the direction normal to the interface leads to
a discrete spectrum of quantized subband states, while motion in
the lateral dimensions is essentially free. In the limit that
electronic excitations take place within only a single subband,
the system can be idealized as strictly
two-dimensional with no spatial extent in the normal direction.
In this situation, the electronic density has the
form $n(x,y,z)=n(x,y)\delta (z)$, where the planar density $n(x,y)$
is defined by whatever additional lateral confining potential
$v_{\rm ext}(x,y)$ is imposed on the system. 

To determine the
equilibrium density distribution $n(x,y)$, we shall make use of
the Thomas-Fermi-Dirac-von Weisz\"acker energy functional:
\begin{eqnarray}
E[n] &=& \int d{\bf r} \left[ C_{1} n^2 + C_{2} \frac{\mid \nabla n(\bf{r}) \mid^{2}}{n(\bf{r})} - C_{3} n^{3/2} \right] \nonumber \\ &+&
\frac{1}{2} \int d{\bf r} \int d{\bf r}' 
\frac{n({\bf r}) n({\bf r}')}{\mid {\bf r} - {\bf r}' \mid} +
\int d{\bf r}\, v_{\rm ext}({\bf r}) n({\bf r})\,.
\label{functional}
\end{eqnarray} 
The first term in (\ref{functional}) is the Thomas-Fermi kinetic 
energy, the second term is the von Weizs\"acker correction to the 
kinetic energy, and the third term is the
Dirac local exchange energy\cite{chaplik71}.   For simplicity, we 
neglect any correlation contribution.
The coefficients in (\ref{functional}) are given by (atomic
units, $e^{2}/\epsilon = m^{\ast} = \hbar = 1$, are used
throughout)

\begin{equation}
C_{1} = \frac{\pi}{2},~~~ C_{2} = \frac{\lambda_{w}}{8},~~~ 
C_{3} = \frac{4}{3}\sqrt{\frac{2}{\pi}}~.
\label{coeff}
\end{equation}
In the von Weizs\"acker coefficient $C_{2}$, the parameter
$\lambda_{w}$ is chosen to have the value $0.25$, which was found in 
other applications\cite{chizmeshya88} to provide the best agreement 
between the TFDW and full density functional theory (DFT) calculations.  
The last two terms in Eq.~(\ref{functional}) are the Hartree 
self-energy of the electrons and the interaction with the external 
potential, respectively.

The equilibrium properties are obtained by finding the variational 
minimum of Eq.~(\ref{functional}). This leads to the Euler-Lagrange 
equation
\begin{equation}
\frac{\delta E[n]}{\delta n({\bf r})} - \mu = 0~,
\label{variation}
\end{equation}
where the Lagrange multiplier (chemical potential) $\mu$
serves to fix the total number of electrons, $N$. By introducing a wave
function according to the prescription
$n({\bf r}) = \psi^{2}({\bf r})$, a straightforward
calculation yields 

\begin{equation}
- \frac{\lambda_{w}}{2} \nabla^{2} \psi ({\bf r}) +
v_{\rm eff} ({\bf r}) \psi ({\bf r}) = \mu \psi ({\bf r})~,
\label{SE}
\end{equation}
where the effective potential is given by

\begin{equation}
v_{\rm eff}({\bf r}) = 2C_{1} \psi^{2}({\bf r}) - 
\frac{3}{2}C_{3}\psi ({\bf r}) + \phi ({\bf r})+v_{\rm ext}({\bf r})~.
\label{veff}
\end{equation}
Here, $\phi({\bf r}) = \int d{\bf r}' n({\bf r}')/|{\bf r} - 
{\bf r}'|$
is the electrostatic potential arising from the electronic 
density $n({\bf r})$.  Eq.~(\ref{SE}) is a nonlinear 
equation in $\psi({\bf r})$, and must be solved self-consistently.  
The required solution to Eq.~(\ref{SE})
is the ground state wave function, hereafter denoted by 
$\psi_{0}$, and the ground state energy eigenvalue defines
$\mu$.

This formulation of the equilibrium properties of the electronic
system is quite general in the sense that the physical situation
is completely specified by the form of the external potential.
For the case of 1D modulation, the potential $v_{\rm ext}(x)$ 
will be assumed to be a
smooth, periodic function of $x$ with period $a$. This potential
can of course be represented as a Fourier
series, but we shall consider only the simplest form 
\begin{equation}
v_{\rm ext}(x) = - V_{0}~{\rm cos}\left(\frac{2 \pi x}{a}\right)
\label{modpot}
\end{equation}
having a single Fourier component.
This choice is sufficient to investigate the important effects of
modulation on the properties of a 2DEG, and
has been applied successfully by other 
workers \cite{wulf90b,dahl90,frank97}
to model laterally microstructured field-effect devices used in
FIR experiments.

The translational invariance of the system along the ${y}$-direction,
together with the periodicity
of the external confining potential, implies that the
solution to Eq.~(\ref{SE}) will only depend on $x$ and will have
the property that
$\psi_{0}(x+a) = \psi_{0}(x)$.  Thus, we can restrict
our calculation to the unit cell $x \in [-a/2,a/2]$. For
potentials such as (\ref{modpot}) with inversion symmetry about the
center of the cell, the desired solutions are those satisfying
the boundary conditions
\begin{equation}
\psi_{0}'\left (\pm \frac{a}{2} \right ) = 0~.
\label{boundary_condition}
\end{equation}
To complete the determination of $\psi_{0}(x)$, we must finally
impose the normalization condition
\begin{equation}
{1\over a} \int_{-a/2}^{a/2} dx \, |\psi_{0}(x)|^2
= \bar n_{2D}\,,
\label{normalization}
\end{equation}
where $\bar n_{2D}$ is the average electron density in the
modulated 2DEG.

Before discussing the nature of the ground state solution to 
Eq.~(\ref{SE}), we first
note some technical points involved in the calculation. As is typical 
of DFT schemes, the solutions are obtained using an 
iterative procedure, and issues of numerical stability 
and convergence therefore arise. The usual source of numerical 
instabilities is the long-ranged nature of  the Coulomb interaction. 
In the course of iterating, a small deviation of the charge 
density from its true value leads to a change in the effective 
potential that tends to induce an overcompensating screening
charge on the
next iteration.  In our particular case, this results in an
oscillation of the charge between the central and outer regions of
the unit cell, which inhibits the desired convergence
of the iterative procedure.  This problem can be remedied 
by a combination of two strategies.
First, we introduce an algebraic mixing of the effective potential

\begin{equation}
v_{\rm eff}^{(i+1)} = 
\alpha~v_{\rm eff}[n^{(i)}]~+~(1 - \alpha)~
v_{\rm eff}^{(i)}\,,
\label{amix}
\end{equation}
where $v_{\rm eff}^{(i)}$ is the effective potential on the
$i$-th iteration, $n^{(i)}$ is the density generated by this
potential and $v_{\rm eff}[n^{(i)}]$  is the effective potential
that this density would lead to. The mixing parameter $\alpha$
lies between zero and one and reducing its value reduces the
change in the potential from one iteration to the next.
This by itself will eliminate the instability in some cases,
as for example when the modulation periods are relatively small.
However, to access the physically relevant modulation periods
($100~{\rm nm} < a < 1500~{\rm nm}$ in GaAs), an extremely small 
value of $\alpha$ would have to be used, which results in an 
unacceptably slow rate of convergence.
This problem can be overcome by means of a second
strategy in which the long-range nature of the Coulomb potential is 
effectively screened. The screening is conveniently formulated
in Fourier space using the algorithm

\begin{equation}
\phi^{(i+1)}({\bf q}) = \frac{2 \pi n^{(i)}({\bf q})}{Q} +
\left (1- \frac{q}{Q} \right ) \phi^{(i)}({\bf q})~,
\label{screenedpot}
\end{equation}
where $\phi({\bf q})$ and $n({\bf q})$ are the 2D Fourier transforms
of the electrostatic potential and 2D electron density, respectively, 
$Q = \sqrt{q^2 + \kappa^2}$, and
${\kappa}$ is an artificial screening parameter.  
One can check that the first term on the right hand side of
(\ref{screenedpot}), in which the density acts as a source,
leads to an exponentially decaying potential in real space.
Nevertheless, at
self-consistency, $\phi^{(i+1)}({\bf q}) = \phi^{(i)}({\bf q}) =
\phi({\bf q}) \equiv 2\pi n({\bf q})/q$, which is just the 
Fourier transform of the actual
Coulomb potential $\phi({\bf r})$.  By choosing
${\kappa}$ of order unity, we were able to obtain convergence in
all cases of interest with $\alpha > 0.005$. Although these
values of $\alpha$ are still rather small, they provided a rate of
convergence which was significantly better than that
achieved with the algebraic mixing scheme alone.

The solution to Eq.~(\ref{SE}) is shown in Fig.~\ref{fig1} for a
range of potential modulation amplitudes. It is useful to
represent these amplitudes in the dimensionless form
$V_{0}/E_{f}$, where $E_f = {1\over 2} k_f^2$ is the Fermi
energy of the uniform 2DEG and the Fermi wavevector $k_f$ is defined
by $k_{f}^{2} \equiv 2 \pi {\overline{n}}_{2D}$.
With an application to GaAs in mind, typical densities are in
the range $10^{11}$--$10^{12} {\rm cm}^{-2}$. Using the
effective mass $m^{\ast} = 0.067 m_{e}$ and static dielectric
constant $\epsilon = 13.0$ of GaAs, the atomic unit of length is 
the effective Bohr radius $a_{0}^{\ast} = 103~$\AA, and the
unit of energy is the
effective Rydberg $Ry^{\ast}=e^{2}/2\epsilon a_{0}^{\ast}=5.4$ meV.
In these units, a density of $10^{11} {\rm cm}^{-2}$ is
approximately $0.1(a_{0}^{\ast})^{-2}$; this is the value
of the average density used to generate the density
distributions in Fig.~\ref{fig1}. The equilibrium densities in
this figure span the range from weak to moderate to strong
modulation, and illustrate the way in which the cell-boundary
density decreases with increasing values of $V_0$. In the
extreme localized limit, the system consists of 
isolated quantum wires, each of which is confined within an 
approximately harmonic potential well of
curvature $2\pi^2 V_0/a^2$. It can be seen that the density in
this limit is similar to the semi-circular profile found in the
classical electrostatic approximation\cite{shikin90}, and only deviates
significantly from this form at the edges of the quantum wire
as a result of the quantum mechanical barrier penetration included
within the TFDW approximation.

In Fig.~\ref{fig2}, we have constructed a `phase' diagram in the
$(a k_{f}$--$V_{0}/E_{f})$ plane which marks
the boundary between localized and delocalized solutions
of Eq.~(\ref{SE}). To generate these curves, we have used 
the criterion
$n_{0}(\pm \frac{a}{2}) = 0.1 {\overline{n}}_{2D}$.  This
condition is rather arbitrary, and a value 
somewhat greater than or less than $1/10$
could have been used to generate a similar set of curves.
The solid and dashed curves correspond to average 2D densities of
${\overline{n}}_{2D}=0.1$ and ${\overline{n}}_{2D}=1.0$,  
respectively and span the range of densities of physical interest.
The open circles represent the `phase' boundary in the case of
noninteracting electrons obtained by ignoring the electrostatic
and exchange potentials in (\ref{veff}). In this limit, one can
show from Eq.~(\ref{veff}) that the scaled wavefunction 
$k_f \psi_0(x)$ as a function
of the scaled length $x/a$ depends on the density only through
the two parameters $a k_{f}$ and $V_{0}/E_{f}$. As a
result, the phase boundary in the $(a k_{f}$--$V_{0}/E_{f})$
plane is a universal curve which is valid for any density. It can be
seen that this
scaling property of the wavefunction does not apply
once interactions are included. We have also indicated by the
labeled points the values of the parameters corresponding to
three of the density profiles in Fig.~\ref{fig1}. These were
obtained for a density of ${\overline{n}}_{2D}=0.1$, so that the
position of the points relative to the solid curve in
Fig.~\ref{fig2} is relevant; point ``(b)" is associated with a
strongly modulated but extended
density while point ``(c)" is well into the localized region.

The parameter $ak_{f}$ can be used to distinguish two regimes of
interest: the rapidly varying regime $ak_{f} \ll 1$ where the
modulation period is small compared to the Fermi wavelength of
the average density, and the slowly varying regime $ak_{f} \gg 1$.
In the former limit, all of the curves approach an
asymptote which is weakly dependent on density. This limit of
strong quantum confinement can be understood simply in terms of
the criterion that the potential barrier height, $V_0$, is large
compared to the quantum mechanical zero point energy
$\hbar^2/ma^2$. We thus conclude that the `phase' boundary is
given approximately by $V_0/E_f \sim (ak_f)^{-2}$, which is
consistent with the numerical calculations. In this limit, the
quantum kinetic energy is deciding the question of localization.
In the opposite limit of a slowly varying potential ($ak_{f} \gg
1$), the von Weizs\"acker kinetic energy becomes negligible and we
recover the Thomas-Fermi (TF) approximation. The experimental 
situations of interest are typically  in this TF limit.
The `phase' boundary now
approaches a straight line with zero slope in the
noninteracting case (open circles) and finite slope in the interacting
case (solid and dashed lines). The behavior of the noninteracting curve
follows from the TF density, $n_{TF}(x) = \bar n_{2D} + {\textstyle{V_0}
\over \textstyle{\pi}} \cos({\textstyle{2\pi x}\over \textstyle{a}})$, 
and our criterion for
localization implies $V_0/E_f \simeq 1$, which is to be expected when
the energy of the gas is exclusively kinetic. On the other hand, the
behavior of the interacting `phase' boundaries can be understood in
electrostatic terms. The external potential $v_{\rm ext}(x)$ can be
viewed as arising from a modulation of the external positive background
density about the average value $\bar n_{2D}$: the amplitude of this
density modulation, $\Delta n_{\rm ext}$, is related to $V_0$ by $V_0 =
{\textstyle{2\pi} \over \textstyle{G}} \Delta n_{\rm ext} = a 
\Delta n_{\rm ext}$. Since the
electrons simply neutralize the positive background locally in the TF
limit, the criterion for localization becomes $V_0 \simeq 
a\bar n_{2D}$, which implies 
$V_0/E_f \simeq {\textstyle{1} \over \textstyle{\pi k_f}} 
(ak_f)$. Thus, the `phase'
boundary has a slope inversely proportional to the square root of the
average density, consistent with the results  in Fig.~\ref{fig2}.

We may also consider general solutions of Eq.~(\ref{SE}) which for 
the case of one-dimensional periodic modulation will take the form 
of Bloch-like states in the $x$ direction, $\psi_{nq_x}
(x) = e^{i q_x x } u_{nq_x}(x)$, where $u_{nq_x}(x)$
is a periodic function of $x$, multiplied by a plane wave factor,
$e^{iq_y y}$, in the
$y$ direction.  Here, $q_x$ is restricted to the first Brioullin zone,
$-\pi/a \le q_x \le \pi/a$, and $n$ is a band index. The ${\bf q} = 0$
state in the lowest band is the ground state $\psi_0(x)$ considered
previously. Although the significance of these general solutions is not
immediately apparent, we shall see that they have some relevance to
the calculation of the collective excitations in a modulated 2DEG,
which similarly exhibit a Bloch-like structure. For the purpose of
comparison with the plasmon bands to be calculated, we show in Fig.~\ref{fig3}
the TFDW energy bands in the $x$ direction for the case of moderate modulation
(``(b)" in Fig.~1).

\section{Collective Excitations}
\label{collective}
\subsection{Hydrodynamic Equations}
\label{hydrodynamics}

In order to determine the plasma modes in the modulated 2DEG, we adopt 
the TFDW hydrodynamic approach developed previously\cite{zaremba94}. This is
based on the usual continuity equation

\begin{equation}
\frac{\partial n}{\partial t} + \nabla \cdot (n {\bf v}) = 0~,
\label{continuity}
\end{equation}
and the momentum equation

\begin{equation}
 n \left[ \frac{\partial {\bf v}}{\partial t} +
{\bf v} \cdot \nabla {\bf v} \right] =
n {\bf F} -  n {\bf v} \times \bbox{\omega}_{c}~,
\label{momentum}
\end{equation}
where ${\bf F} = {\bf F}_{\rm int}+ {\bf F}_{\rm ext}$
includes both the internal force acting on the electrons,

\begin{eqnarray}
{\bf F}_{\rm int}({\bf r}, t) = - \nabla \left[ v_{\rm eff}({\bf r}, t) 
- \frac{\lambda_{w }}{2}
\frac{\nabla^{2} \psi ({\bf r},t)}{\psi ({\bf r},t)} \right]\,,
\label{force}
\end{eqnarray}
as well as any additional time-dependent external force, ${\bf F}_{\rm
ext}({\bf r},t)$. The potential term $v_{\rm eff}$ in Eq.~(\ref{force})
contributes the expected force corresponding to the
internal TF pressure and Coulomb--derived potentials, while the
remaining term is associated with the von Weizs\"acker kinetic energy.
For completeness, we
have also included in Eq.~(\ref{momentum}) the magnetic force due to an
externally applied magnetic field, {\bf B}, which we shall take to be 
uniform and
perpendicular to the 2DEG ({\em i.e.,} ${\bf B} = B {\bf \hat z}$).
This force is expressed in terms of the cyclotron frequency
vector $\bbox{\omega}_{c} = e {\bf B }/ m^{\ast} c$.
Although we defer an explicit treatment of  magnetoplasma modes 
to a future paper, we shall develop the equations for this more general
situation as very little additional effort is needed to do so.

Self--consistent solutions to our hydrodynamic equations are obtained by
linearizing the density in
small deviations from equilibrium, {\em viz.,} $n \rightarrow n_{0} +
\delta n$.
If the time-dependent density is represented as $n = \psi^2$,
the fluctuation of the von Weizs\"acker wave function, $\delta
\psi$, is related to  the density fluctuation
by $\delta n = 2 \psi_{0} \delta \psi$.  Retaining only first
order quantities (the velocity ${\bf v}$ is itself first order),
Eqs.~(\ref{continuity}) and (\ref{momentum}) yield

\begin{equation}
\frac{\partial \delta n}{\partial t} + \nabla \cdot (n_{0} {\bf v}) = 0~,
\label{lincont}
\end{equation}
and
\begin{equation}
\frac{\partial {\bf v}}{\partial t} =
\delta {\bf F} - {\bf v} \times \bbox{\omega}_{c}~,
\label{linmom}
\end{equation}
where the fluctuating force is given by

\begin{equation}
\delta {\bf F} = - \nabla \left[ \delta v_{\rm eff} -
\frac{\lambda_{w}}{2 \psi_{0}} \nabla^{2} \delta \psi +
\frac{\lambda_{w}}{2} \frac{\nabla^{2} \psi_{0}}{\psi_{0}^{2}}
\delta \psi \right] + \delta {\bf F}^{\rm ext}~,
\label{fluctforce}
\end{equation}
with 

\begin{equation}
\delta v_{\rm eff} = 4C_{1}\psi_{0}\delta \psi -
\frac{3}{2}C_{3}\delta \psi + \delta \phi~.
\label{flucveff}
\end{equation}
The bracketed term in Eq.~(\ref{fluctforce}) will be denoted by $f$, so
that $\delta {\bf F}^{\rm int} = - \nabla f$.

For the purpose of determining the normal mode frequencies of the system
and the associated mode densities, $\delta {\bf F}^{\rm ext}$ can be 
set to zero. We shall later consider the response of the 
system to external fields in the calculation of the power absorption. 
Due to the translational invariance in the
$y$-direction, all of the fluctuating variables ($\delta n$, $\delta
\psi$, $f$ and ${\bf v}$) will have the form of a propagating
wave $e^{i({q_{y}} y -
\omega t)}$ with $x$-dependent amplitudes. Making use of this
dependence, Eqs.~(\ref{lincont}) and (\ref{linmom}) can be 
expressed in the form

\begin{equation}
-i \omega \delta n + i n_{0} q_{y} v_{y} + \frac{\partial}{\partial x} 
(n_{0} v_{x}) = 0
\label{harmcont}
\end{equation}
and
\begin{equation}
(\omega^{2} - \omega_{c}^{2}) {\bf v} = i\omega \delta 
{\bf F} - (\bbox{\omega}_{c} \times \delta {\bf F})~,
\label{harmmom}
\end{equation}
respectively. In these equations, only the $x$-dependent amplitudes are
displayed. Recalling that $\delta {\bf F} = - \nabla f$, we have

\begin{eqnarray}
{\bbox{\omega}}_{c} \times \delta {\bf F} = (i q_{y} 
\omega_{c} f){\bf {\hat x}} -
\left(\omega_{c} \frac{\partial f}{\partial x} 
\right){\bf {\hat y}}~.
\label{crossproduct}
\end{eqnarray}
The use of Eq.~(\ref{crossproduct}) in Eq.~(\ref{harmmom}), along with
Eq.~(\ref{harmcont}), then leads to

\begin{equation}
\omega(\omega^{2} - \omega_{c}^{2}) \delta n =
q_{y}^{2} \omega n_{0}f 
- \omega \frac{\partial}{\partial x}
\left(n_{0} \frac{\partial f}{\partial x} \right)
-  \omega_{c} q_{y} \frac{\partial n_{0}}{\partial x} f~.
\label{master}
\end{equation}

Eq.~(\ref{master}) can also be expressed in terms of the 
fluctuating wave function by using $n_{0}=\psi_{0}^{2}$ and
$\delta n = 2 \psi_{0} \delta \psi$.  With this substitution, we find

\begin{equation}
\omega (\omega^{2} - \omega_{c}^{2}) \delta \psi = 
\frac{1}{2} q_{y}^{2} \omega \psi_{0} f
- \omega \psi_{0}' f' - \frac{1}{2} \omega \psi_{0} f''
- \omega_{c} q_{y} \psi_{0}' f~,
\label{step1}
\end{equation}
where derivatives with respect to $x$ are now denoted by primes.
Noting that $\psi_{0} f'' = (\psi_{0} f)'' - \psi_{0}'' f - 2 \psi_{0}' f'$,
Eq.~(\ref{step1}) yields,

\begin{eqnarray}
\omega (\omega^{2} - \omega_{c}^{2}) \delta \psi &=&
\frac{1}{2} q_{y}^{2} \omega (\psi_{0} f) +
\frac{1}{2} \omega \frac{\psi_{0}''}{\psi_0} (\psi_0 f) \nonumber \\ 
&-& \frac{1}{2} \omega (\psi_{0} f)''
- \omega_{c} q_{y} \frac{\psi_{0}'}{\psi_{0}} (\psi_{0} f)~.
\label{step2}
\end{eqnarray}
The advantage of this form is that the function $(\psi_{0} f)$ and its
derivatives now appear on the right hand side, where 

\begin{eqnarray} 
\psi_{0} f &=& \psi_{0} \delta v_{\rm eff} + \frac{\lambda_{w}}{2}
\left (\frac{\psi_{0}''}{\psi_{0}} + q_{y}^{2} \right )\delta \psi -
\frac{\lambda_{w}}{2} \delta \psi''\nonumber\\
&\equiv&\hat M \delta \psi + \hat h \delta \psi\,.
\label{psi0f}
\end{eqnarray}
The operator $\hat M$ is defined by ${\hat M} \delta \psi
\equiv \psi_{0} \delta v_{\rm eff}$ and
\begin{equation}
{\hat h} = -\frac{\lambda_{w}}{2} \left ( {d^2 \over dx^2} -
q_y^2 \right ) + v_{\rm eff} - \mu
\label{hhat}
\end{equation}
is just the Hamiltonian (for $q_y = 0$) determining the
ground state von Weizs\"acker wave function $\psi_0(x)$.
Substituting Eq.~(\ref{psi0f}) into (\ref{step2}) yields

\begin{equation}
\omega (\omega^{2} - \omega_{c}^{2}) \delta \psi =
\omega \lambda_w^{-1} {\hat h}({\hat h} + {\hat M})
\delta \psi - \omega_c q_y  \frac{\psi_{0}'}{\psi_{0}} ( {\hat h}+
{\hat M}) \delta \psi\,.
\label{altstep2}
\end{equation}
In the limit of vanishing magnetic field, this reduces to
\begin{equation}
\lambda_w \omega^{2} \delta \psi = {\hat h}({\hat h} + {\hat M})
\delta \psi\,.
\label{zeroB}
\end{equation}
 
Due to the periodicity induced by the modulating potential along
the $x$-direction, the fluctuating part of the wave function
will have the Bloch-like form

\begin{equation}
\delta \psi = e^{i q_{x} x}~\sum_{G} c_{G}
\varphi_{G}~,
\label{fourierforce}
\end{equation}
where $c_{G}$ is a Fourier expansion coefficient and
$\varphi_{G} = \frac{\textstyle{1}}{\sqrt{\textstyle{a}}} e^{i G x}$.  
These basis functions have been chosen to
satisfy the orthonormality condition

\begin{equation}
\int_{-a/2}^{a/2}~dx \varphi^{\ast}_{G}(x)\varphi_{G'}(x) = \delta_{GG'}~,
\label{orthonormal}
\end{equation}
where $G = \frac{\textstyle{2 \pi n}}{\textstyle{a}}$ $(n = 0, \pm 1, \pm2 , ...)$
is a one--dimensional reciprocal lattice vector in the
$x$--direction.  Substituting the Fourier expansion 
(\ref{fourierforce}) into (\ref{altstep2}), we obtain the equation
\begin{eqnarray}
\omega (\omega^{2} - \omega^{2}_{c}) c_{G} =
-\omega_{c} q_{y} \sum_{G'G''} A_{GG'} {\tilde M}_{G'G''} c_{G''} + \omega
\sum_{G'G''} B_{GG'} {\tilde M}_{G'G''} c_{G''}~,
\label{matrixeqn}
\end{eqnarray}
where the various matrices appearing in this equation are defined
as
\begin{eqnarray}
A_{GG'} &=& \frac{i}{\sqrt{a}}(G-G')\overline{\ln \psi_{0}}[G-G']~,
\label{amat}\\
B_{GG'} &\equiv& \lambda_w^{-1} h_{GG'}\nonumber \\
&=& \frac{1}{2} \left[(q_{x} + G)^{2} + q_{y}^{2} \right]
\delta_{GG'} + \frac{1}{\lambda_{w}\sqrt{a}}
\overline{(v_{\rm eff} - \mu)}[G-G']~,
\label{bmat}
\end{eqnarray}
and
\begin{eqnarray}
\tilde M_{GG'} = M_{GG'} + \lambda_w B_{GG'}\,.
\label{Mmatrix}
\end{eqnarray}
In obtaining (\ref{amat}) we have used the identity
$\frac{\textstyle{\psi_{0}'}}{\textstyle{\psi_{0}}} = 
\frac{\textstyle{d}}{\textstyle{dx}}(\ln \psi_{0})$, and
denote the Fourier transform of a periodic function
with an overline, {\em e.g.} 
$\psi_{0}(x) = \sum_{G} {\overline{\psi_{0}}}[G] \varphi_{G}$.

The remaining matrix $M_{GG'}$ in (\ref{Mmatrix}) consists of
several terms. From Eq.~(\ref{psi0f}), we have
\begin{eqnarray}
\hat M \delta \psi &=& \psi_{0} \delta v_{\rm eff} \nonumber \\
&=& 4C_{1} \psi_{0}^{2} \delta \psi - \frac{3}{2} C_{3} \psi_{0} 
\delta \psi + \psi_{0} \delta \phi\,.
\label{step7}
\end{eqnarray}
The last term in (\ref{step7}) involves the $x$-dependent part
of the total electric potential fluctuation, $\delta \phi$.
In terms of the density fluctuation $\delta n$, we have
\begin{eqnarray}
\delta \phi (\bf{r}) &=& \int~d{\bf r}'~\frac{\delta n({\bf r}')}
{\mid {\bf r} - {\bf r}' \mid } \nonumber \\
&=&
\int~d{\bf r}'~
\frac{2 \psi_{0}(x') e^{i(q_{x}x' + q_{y}y')} \sum_{G'}c_{G'}
\varphi_{G'}(x')} {\mid {\bf r} - {\bf r}' \mid } \nonumber \\
&=&
e^{i(q_{x}x + q_{y}y)}\frac{4 \pi}{a} \sum_{GG'}
\frac{e^{iGx}}{\sqrt{(q_{x}+G)^{2}+q_{y}^{2}}}
{\overline{\psi_{0}}}[G-G']c_{G'}~,
\label{deltaphi}
\end{eqnarray}
where we have noted that the 2D Fourier transform of ${r}^{-1}$ is
$(2 \pi /{q})$.  It is clear from this expression that 
$\delta \phi (\bf{r})$ is proportional to $e^{iq_{y}y}$ and that 
the $x$-dependent part has the expected Bloch-like form.
With this result, (\ref{step7}) yields
\begin{equation}
M_{GG'} = M^{K}_{GG'} + M^{X}_{GG'} + M^{H}_{GG'}~,
\label{mtmat}
\end{equation}
where
\begin{equation}
M_{GG'}^{K} = \frac{4 C_{1}}{\sqrt{a}} \overline{\psi_{0}^{2}}[G-G']~,
\label{kinetic}
\end{equation}
\begin{equation}
M_{GG'}^{X} = -\frac{3C_{3}}{2\sqrt{a}} \overline{\psi_{0}}[G-G']~,
\label{exchange}
\end{equation}
\begin{equation}
M_{GG'}^{H} = \frac{4 \pi}{{a}} \sum_{G''}
\frac{\overline{\psi_{0}}[G - G'']~\overline{\psi_{0}}[G'' - G']}
{\sqrt{(q_{x} + G'')^{2} + q_{y}^{2}}}~.
\label{hartree}
\end{equation}
The superscripts $K$, $X$ and $H$ refer respectively to the kinetic, 
exchange and Hartree terms in $\psi_{0}\delta v_{\rm eff}$. It
is clear that the kinetic and exchange matrices depend on
$G-G'$, but the Hartree matrix, which is associated with a
nonlocal operator in position space, has the property
$M_{G+G_0,G'+G_0}^{H}(q_x) = M_{GG'}^{H}(q_x+G_0)$.

Eq.~(\ref{matrixeqn}) is a nonlinear eigenvalue problem which
must be solved numerically for the
eigenvalue $\omega$ and eigenvector $\vec c$.
In principle, the dimension of the eigenvalue problem is
infinite, and for practical purposes, a truncation of the expansion 
to some finite number of $G$ vectors is required.  However, one can 
always check that the results for the modes of interest have 
converged by systematically increasing the number of $G$ vectors.  
Once the eigenvalues $\omega$ have been
determined, the corresponding eigenvectors $\vec c$ can be substituted
into Eq.~(\ref{fourierforce}) to determine the mode densities.

To close this section, we note some general consequences of the
equations we have obtained. In the $q_y \to 0$ limit, (\ref{matrixeqn}) 
reduces to 
\begin{equation}
(\omega^{2} - \omega^{2}_{c}) c_{G} =
\sum_{G'G''} B_{GG'} {\tilde M}_{G'G''} c_{G''}~,
\label{eqnqy0}
\end{equation}
and the magnetic field appears explicitly only on the left
hand side of the equation. This implies that the magnetoplasma
frequencies have the property
\begin{equation}
\omega_n^2(q_x,q_y = 0; B) = \omega_n^2(q_x,q_y = 0; B =0) +
\omega_c^2\,,
\label{magfreq}
\end{equation}
that is, a simple cyclotron shift of the zero-field frequencies,
$\omega_n(q_x,q_y = 0; B =0)$. In addition, the mode densities
in this limit have exactly the same spatial distribution as the
corresponding $B=0$ mode densities.

Another general property can be deduced from Eq.~(\ref{zeroB}) by
making use of the Bloch-state basis, $\psi_{nq_x}(x)$,
introduced at the end of Sec.~II. These states are eigenstates
of $\hat h$ with eigenvalues $\varepsilon_n({\bf q}) = E_n(q_x)
+ {\textstyle{1}\over \textstyle{2}}\lambda_w q_y^2 - \mu$, where
$E_n(q_x)$ are the 1D band energies and the chemical potential
$\mu$ is equal to  $E_0(0)$, the ${\bf q} = 0$ energy of the
lowest band ($n=0$). Expanding $\delta \psi$ as
\begin{equation}
\delta \psi = \sum_n d_n \psi_{nq_x}\,,
\label{alternate}
\end{equation}
and substituting this expansion into Eq.~(\ref{zeroB}), we obtain
the eigenvalue problem in the alternate form
\begin{equation}
\lambda_w \omega^2 d_n = \varepsilon_n({\bf q}) \sum_{n'} \left
[ \varepsilon_n({\bf q}) \delta_{nn'} + M_{nn'} \right ]
d_{n'}\,.
\label{alteqn}
\end{equation}
By its definition, $\varepsilon_0(0) = 0$ for the lowest
band, and as a result, Eq.~(\ref{alteqn}) will have a
nontrivial solution at ${\bf q} =0$ with frequency $\omega
= 0$. In other words, the lowest plasmon band disperses from
zero at ${\bf q} =0$ and, according to Eq.~(\ref{magfreq}), this
implies that there will be a magnetoplasmon branch which
disperses from $\omega_c$.

\subsection{Power Absorption}
\label{power_absorption}

The main method for studying the collective modes in these systems is by
means of FIR absorption experiments\cite{gudmundsson95,mackens84,hansen87,brinkop88,demel88,drexler92,frank97}.  
To make contact with these
experiments, we consider in this section
the calculation of the power absorption which, in addition to
the mode frequencies, contains information about the oscillator
strengths of the observed excitations.

The instantaneous power absorption is given by

\begin{equation}
P(t) = \int d{\bf r}~{\bf j}^{\rm ind}({\bf r},t)\cdot
{\bf E}^{\rm ext}({\bf r},t)~,
\label{powerabs}
\end{equation}
where ${\bf j}^{\rm ind}({\bf r},t)$ is the current induced by an
external electromagnetic field. For a uniform radiation field
polarized in the $x$--direction and 
incident normally on the sample, the external field is spatially
uniform and is given by ${\bf E}^{\rm ext} ({\bf r}, t) 
=\frac{1}{2}E_{0}(e^{-i\omega t} + e^{i \omega t}) {\bf \hat x}$.
The physically relevant quantity is the time-averaged power
absorption which is given by

\begin{equation}
\langle P \rangle_{t} = \frac{1}{2} 
E_{0}\int d{\bf r}~{\rm Re}~j_x^{\rm ind} ({\bf r},\omega)~.
\label{avgpwr}
\end{equation}
Since the induced current $j_{x}^{\rm ind}$ for the situation of
interest is a periodic function of $x$ and independent of $y$,
the time-averaged power absorption per unit area is given by
\begin{equation}
{\langle P \rangle_{t} \over A} = \frac{1}{2\sqrt{a}} E_{0}~{\rm Re}~
{\overline{j_{x}^{\rm ind}}}[ G=0,\omega]~,
\label{pwr_g}
\end{equation}
where
\begin{equation}
{\overline{j_{x}^{\rm ind}}}[G,\omega] = \frac{1}{\sqrt{a}} 
\int_{-a/2}^{a/2} dx~ e^{-iGx} j_{x}^{\rm ind}(x,\omega)
\label{jg}
\end{equation}
is the Fourier coefficient of the induced current.

The current density is determined by $j_x^{\rm ind}(x,\omega)
= -n_0(x)v_x(x,\omega)$ where the velocity is the solution of
\begin{equation}
\frac{\partial {\bf v}}{\partial t} =
- \gamma {\bf v} + \delta {\bf F} -
{\bf v} \times \bbox{\omega}_c~.
\end{equation}
We have included in this equation a phenomenological relaxation
rate, $\gamma$, which accounts for momentum
nonconserving scattering processes of the electrons.
Because of the relaxation rate, the frequency $\omega$ appearing in
Eq.~(\ref{harmmom}) is now replaced by
$\tilde{\omega} = \omega + i\gamma$.
Noting that $\delta {\bf F}$ now includes the additional term
$\delta F_{x}^{\rm ext} = - E_{0}$, we obtain the following 
expression for the current density at ${\bf q} =0$,

\begin{equation}
j_{x}^{\rm ind}(x,\omega) = -n_{0}(x) v_{x}(x,\omega) = 
\frac{i \tilde{\omega}}{\tilde{\omega}^{2} - \omega_{c}^{2}} n_{0}
\left( \frac{\partial f}{\partial x} + E_{0} \right)~.
\label{j_xtemp}
\end{equation}
Taking the ${\bf q} = 0$ limit of the continuity equation in 
(\ref{harmcont}), along with Eq.~(\ref{j_xtemp}), we find

\begin{equation}
\omega (\tilde{\omega}^{2} - \omega_{c}^{2}) \delta n =
- {\tilde{\omega}} \frac{\partial}{\partial x} \left(n_{0} 
\frac{\partial f}{\partial x}
\right) - {\tilde{\omega} E_{0}} \frac{\partial n_{0}}{\partial x}~.
\label{augmom}
\end{equation}
Thus, the net effect of including the external driving 
field ${\bf E}^{\rm ext}$ is to convert the eigenvalue problem in 
Eq.~(\ref{matrixeqn}), into the set of inhomogeneous equations

\begin{equation}
\omega (\tilde{\omega}^{2} - \omega_{c}^{2}) c_{G} -
\tilde{\omega} \sum_{G'G''} B_{GG'} \tilde{M}_{G'G''} c_{G''} =
-i \tilde{\omega} G \overline{\psi_{0}}[G] E_{0}~.
\label{inhomo}
\end{equation}

Substituting Eq.~(\ref{j_xtemp}) into Eq.~(\ref{jg}), we obtain

\begin{equation}
{\overline{j_{x}^{\rm ind}}}[G=0,\omega] = 
\frac{-2 \tilde{\omega}}{\sqrt{a}(\tilde{\omega}^{2} - \omega_{c}^{2})}
\sum_{G} G~\overline{\psi_{0}}[G] f_{G} +
\frac{i \tilde{\omega}E_{0}}{\tilde{\omega}^{2} - \omega_{c}^{2}}
\overline{n_{0}}[G=0]~,
\label{jxG0}
\end{equation}
where, in deriving this result, we have used the fact 
that $(n_{0}f)$ is periodic, $n_0 = \psi_0^2$, and 
$\overline{\psi_0}[-G] = \overline{\psi_0}[G]$.
The quantity $f_G$ is the Fourier coefficient of $(\psi_0 f)$ defined
in Eq.~(\ref{psi0f}), 
and is related to $c_G$ by the equation
\begin{equation}
f_{G} = \sum_{G'} \tilde{M}_{GG'} c_{G'}~.
\end{equation}
The substitution of (\ref{jxG0}) into
(\ref{pwr_g}) provides our final expression  for
the power absorption. Since $\overline{n_{0}}[G=0] = \sqrt{a}\bar
n_{2D}$, the last term in Eq.~(\ref{jxG0}) contributes 
an absorption peak at the cyclotron frequency, $\omega_c$, as
found for a uniform 2DEG. This peak shifts to 
$\omega = 0$ for $B = 0$, and represents the expected Drude-like
absorption associated with the resistive losses in the 2DEG. The
other term on the right hand side of Eq.~(\ref{jxG0}) accounts
for the density inhomogeneity and has the effect of reducing the
amount of Drude absorption.

The cancellation of the Drude peak can be seen more clearly
by writing the power absorption in an alternate form. Starting 
with the continuity equation (\ref{harmcont}) with ${\bf q} = 0$
\begin{equation}
i \omega \delta n + \frac{\partial}{\partial x} j_x^{{\rm ind}} = 0~,
\end{equation}
multiplying by $x$ and integrating over a unit cell, we find
\begin{equation}
{\overline{j_{x}^{\rm ind}}}[G=0,\omega] = \frac{i \omega}{\sqrt{a}}
\int_{-a/2}^{a/2} x \delta n(x)~dx + \sqrt{a} j^{\rm ind}(a/2)~.
\label{jxG0_2}
\end{equation}
The substitution of this expression into Eq.~(\ref{pwr_g}) then
yields 
\begin{equation}
{\langle P \rangle_{t} \over A} = -\frac{1}{2} \omega E_{0}~{\rm Im}~
{1\over a} \int_{-a/2}^{a/2}\,x\delta n(x,\omega)\,dx + {1\over
2} E_0~{\rm Re} j_{x}^{\rm ind}(x= a/2,\omega)~.
\label{altexp}
\end{equation}
In the localized limit, the
boundary current $j_{x}^{\rm ind}(x= a/2,\omega)$ vanishes and
the power absorption is then determined by the 
induced dipole moment of the charge density fluctuation. In this 
case, there is no
absorption at the cyclotron frequency, $\omega_c$, despite the
appearance of the resonant denominators in Eq.~(\ref{jxG0}).
In the opposite limit of a weakly modulated system, the induced
dipole moment will be small and the power absorption will be 
dominated by the second term in (\ref{jxG0_2}). We then recover
the Drude absorption discussed previously.

\section{The 2D to 1D Crossover}
\label{2dto1d}

In this section, we study the plasma modes of a 2DEG subjected
to the modulating potential given in Eq.~(\ref{modpot}).
Our main interest is in the evolution of these modes as a
function of the strength of the modulation. The crossover from
2D to 1D behaviour will occur in the vicinity of the `phase'
boundary illustrated in Fig.~1. Once the modulation is
sufficiently strong, the electron layer separates into an array
of quantum wires and we can expect very different behaviour from
the original 2D situation. Although the theory in Sec.~III was
developed with the inclusion of a magnetic field, we shall
restrict our investigation to the zero field limit, in which
case Eq.~(\ref{matrixeqn}) reduces to
\begin{equation}
\omega^{2} c_{G} =
\sum_{G',G''} B_{GG'} {\tilde M}_{G'G''} c_{G''}~.
\label{weak_B=0}
\end{equation}
All the mode frequencies presented in this section are based on
an analysis of this equation.

\subsection{The Uniform 2DEG}
\label{uniform}

It is useful for the purpose of orientation to begin with the
homogeneous 2DEG ($v_{\rm ext} =0$) as treated in the TFDW
approximation. The interaction matrices in 
Eqs.~(\ref{kinetic})-(\ref{hartree}), as well
as the $B$-matrix in (\ref{bmat}), are diagonal in this case and we
readily obtain from (\ref{weak_B=0}) the plasmon dispersion relation

\begin{eqnarray}
\omega^{2}_0(q) &=& 2 \pi {\overline{n}_{2D}} q +
\left ( 2C_{1}{\overline{n}_{2D}} -
\frac{3}{4}C_{3}\sqrt{{\overline{n}_{2D}}}\right ) q^{2} +
\frac{\lambda_{w}}{4}q^{4}\nonumber \\
&=& k_f^2 q + \left ( {1\over 2}k_f^2 -{1\over \pi} k_f \right )
q^2 +\frac{\lambda_{w}}{4}q^{4}\,.
\label{dispersion_zero}
\end{eqnarray}
At long wavelengths, this gives the expected 2D plasmon frequency
$\omega_{2D} = \sqrt{2\pi \bar n_{2D} q}$ which has the
characteristic $\sqrt{q}$ dependence. At shorter wavelengths,
the effects of the TF ($C_1$), exchange ($C_3$) and von
Weizs\"acker ($\lambda_w$) energies become of increasing
importance. It is interesting to note that the exchange
interaction gives a negative $q^2$ coefficient which counteracts
the positive dispersion coming from the TF kinetic energy. As a
result, the $q^2$ coefficient goes to zero at a density having
$k_f = 2/\pi$.

\subsection{Weak Modulation}
\label{weak}

We now consider a weak modulating potential corresponding to the
phase-point ``(a)'' in Fig.~\ref{fig2}, with coordinates $(a
k_{f} = 31.71, V_{0}/E_{f} = 6.4)$.  
In Fig.~\ref{fig4} we illustrate the $q_x$-dispersion for $q_{y}=0$,
as determined from Eq.~(\ref{weak_B=0}). Also shown for
comparison as the solid line is the uniform 2DEG dispersion
relation given by Eq.~(\ref{dispersion_zero}). The similarity of the
dispersion curves for the two cases is notable. The only obvious
difference is the appearance of a gap in the plasmon dispersion
at the zone boundary, $q_x = \pi/a$, which is associated with the
development of plasmon bands. This problem was treated
theoretically by Krasheninnikov and Chaplik\cite{krasheninnikov81} 
using degenerate perturbation theory. They found that
the size of the gaps induced by the density modulation is given by 
$\Delta \sim (\vert N_{n}\vert /\bar n_{2D})\omega_{n}$, where
$\omega_n$ is the unperturbed plasma frequency for wavevector
$q_x \equiv q_{n} = n \pi/a$ and $N_{n}$ is the $n$th coefficient 
in the Fourier expansion of the 2D electron density ($N_n =
\overline{n_0}[2\pi n/a]/\sqrt{a}$ in our notation; odd values
of $n$ give rise to zone boundary (ZB) gaps, while even values
give rise to zone center (ZC) gaps). Since their
analysis is based on a different formulation of the problem and
includes only the Hartree interaction, we present for completeness 
a perturbation theory calculation of the gaps within the
TFDW hydrodynamics in Appendix \ref{perturbation}. We find that 
the gaps are determined by the equation
\begin{equation}
\omega_{\pm}^{2} = \omega_0^2(q_n) \pm \left ( \omega_0^2(q_n) -
{3\over 8} C_3 \sqrt{\bar n_{2D}} q_n^2 +{3\over 4} \lambda_w
q_n^4 \right ) {\overline {n_0}[2\pi n/a] \overwithdelims || 
\overline {n_0}[0]}\,,
\label{gap}
\end{equation}
which is consistent with the result of Krasheninnikov 
and Chaplik\cite{krasheninnikov81} in the long wavelength
limit where the Hartree interaction is dominant. In Fig.~\ref{fig15} 
of Appendix \ref{perturbation}, we
show some of the Fourier coefficients $\overline{n_0}[2\pi n/a]$ 
as a function of $V_0/E_f$. The $\overline{n_0}[2\pi/a]$  
coefficient is linear in $V_0$ whereas the
higher Fourier components are at least of order $V_0^2$ and are
therefore much smaller in the weakly modulated regime. This
accounts for the relatively small magnitude of the gaps in 
Fig.~\ref{fig4} as compared to the lowest zone boundary gap.

In Fig.~\ref{fig5}, we show the plasmon dispersion in the $y$ direction
({\em i.e.}, perpendicular to the modulation direction) for modes at 
the zone center ($q_x = 0$, solid circles) and at the zone boundary
($q_x = \pi/a$, open circles). Specific modes of interest that
will be referred to in the text are indicated by encircled numbers.
Of the $q_x=0$ modes, there is one which starts at zero frequency
(mode ``1"), 
as explained at the end of Sec.~\ref{hydrodynamics}. 
To determine the small-$q_y$ behaviour analytically, we make use
of the long--wavelength form of $\tilde{M}_{GG'}$ derived in
Appendix B. It is shown there that the 2D 
plasmon dispersion for $q_x =0$ with $q_y \rightarrow 0$
is given by $\omega^2 \simeq 2\pi \bar n_{2D} q_y$, 
the usual plasma frequency of a uniform 2DEG. This is a
general result independent of the degree of modulation of the
2DEG. In other words, the modulation of the 2DEG has no effect
on the long--wavelength dispersion of plasmons in this direction. 
However, as discussed in more detail in Sec.~\ref{moderate},
this is not the case for plasmons propagating in the $x$
direction.

There are several other notable features in Fig.~\ref{fig5}. 
The first concerns the modes at $q_x = \pi/a$ which behave as
$\delta n(x+a) = -\delta n(x)$ and have a periodicity of $2a$. 
The two lowest modes (the ``2-8" and ``3-7" branches)
indicated by open circles in Fig.~\ref{fig5} 
cross close to $q_y = 1$. This implies that the gap at the zone
boundary first decreases to zero as a function of $q_y$, and then
increases again. Such a crossing is possible since the two
modes in question have different symmetries. 
To illustrate this more clearly, we show the induced charge
density fluctuations for various modes in Fig.~\ref{fig6}.
The left panel of this figure corresponds to $q_y =0$ while the right
panel is for $q_y = 3\pi/a$.  The sequence from top to bottom 
illustrates the five lowest modes, starting with the $q_x = 0$
mode in the top panel, followed by the two $q_x = \pi/a$ modes in the
middle panel and finally the next pair of $q_x = 0$ modes in the
bottom panel. The right panel of Fig.~\ref{fig6} shows how these
mode densities change when $q_y$ is increased to $3\pi/a$.
It is the pair of $q_x = \pi/a$ modes in the middle panel which
exhibits the frequency crossing in Fig.~\ref{fig5}. For $q_y$
values up to the crossing point, the lower frequency mode 
(the ``2-8" branch) is
symmetric with respect to reflections about $x=0$, while the
next highest mode 
(the ``3-7" branch) is antisymmetric. The ordering in frequency of
these two modes is then reversed at the crossing point, with the
odd-parity mode lying lower. Similar behaviour was found for the
model densities with weak modulation considered in 
Ref.~[\onlinecite{eliasson86b}], but for strong modulation 
an anticrossing behaviour was observed. As we shall see, 
the crossing behaviour we find persists for even stronger 
modulations (see Fig.~\ref{fig9}).

A further examination of Fig.~\ref{fig5} shows the odd-parity
mode at $q_x = \pi/a$ approaching the lowest--lying $q_x = 0$
mode with increasing $q_y$ (points ``6" and ``7"). 
The reason for this can be explained 
by comparing the mode density (``6") in the top-left panel of
Fig.~\ref{fig6} with the odd-parity density (``7") in the middle-right
panel. Both mode densities are seen to be localized at the cell
boundaries where the equilibrium density is
lowest, and both have a very similar spatial profile in this
region. The near degeneracy of these modes for large $q_y$
indicates that there is a weak interaction between the density 
fluctuations localized at adjacent cell boundaries and that as a
result, these modes effectively propagate independently of each 
other. This behaviour is analogous to the situation in a
metallic slab where the symmetric and antisymmetric surface 
plasmons become degenerate at large wave vectors. 
The localization of the mode density at the cell
boundaries increases with increasing $q_y$, implying that the
modes are channeled in their propagation along the low density
part of the gas.

We also see a similar convergence of the other $q_x = \pi/a$
mode with the next higher $q_x = 0$ mode at large $q_y$. The
explanation for this is the same as given above. The density
fluctuation ``8" shown by the solid line in the right-middle panel 
of Fig.~\ref{fig6} is the even-parity version of the odd-parity
density fluctuation ``9" shown by the dashed curve in the lower-right
panel. A similar pairing of $q_x = 0$ and $q_x = \pi/a$ modes would
also be expected for the higher lying modes in Fig.~\ref{fig5}.
This behaviour is in fact more evident when the modulation
amplitude of the equilibrium density increases (see Figs.~\ref{fig9}
and \ref{fig12}). We should emphasize that the convergence of 
pairs of $q_x = 0$
and $q_x = \pi/a$ mode frequencies implies that these modes
exhibit a very weak dispersion with respect to $q_x$. The
absence of an effect of the $e^{iq_xx}$ phase modulation from one 
unit cell to the next reflects the lack of interaction between 
adjacent density fluctuations.

Returning to Fig.~\ref{fig4}, we see that the gaps at the zone center 
are unobservably small for this case of weak modulation. However,
each of the higher lying modes at $q_x = 0$ corresponds to two
distinct modes. This becomes apparent for the first excited $q_x
= 0$ modes in Fig.~\ref{fig5} (starting at the point labeled
``4" and ``5")
where it is seen that the frequencies 
separate with increasing $q_y$. The lower-left
panel in Fig.~\ref{fig6} shows the mode densities for this pair
of modes at $q_y = 0$, and the lower-right panel shows the 
densities at $q_y = 3\pi/a$. One mode has even parity with
respect to the center of the cell, and the other has odd parity.
The latter mode eventually evolves with increasing modulation
into the lowest odd-parity
mode of an isolated wire. Since this mode has a finite dipole
moment, it will couple to an external radiation field and will
contribute to the power absorption as indicated by Eq.~(\ref{altexp}). 
To illustrate this, we show in Fig.~\ref{fig7} the calculated
power absorption for different modulations of the 2DEG. The
lowest curve labeled ``(a)'' corresponds to the case of weak 
modulation being considered here and shows a small peak at the
frequency of the first excited $q_x = 0$ mode ``4". This is
essentially a bulk 2D plasmon at a wavevector $q=2\pi/a$.
In principle,
other $q_x = 0$ odd-parity modes should be observable, but
their oscillator strengths are too small to show up
in the power absorption. In fact, the dominant feature in the
power absorption of curve ``(a)" is the strong Drude peak at
$\omega = 0$ which is to be expected since the system is only
weakly perturbed from a homogeneous 2DEG. We shall return to a
more systematic discussion of the power absorption later.

\subsection{Moderate Modulation}
\label{moderate}

As the modulation is increased, the 
equilibrium density profile becomes more localized 
(see the curve labeled by ``(b)'' in Fig.~\ref{fig1}). The
effect of this increased modulation on the wavevector dispersion
along the $x$-direction is illustrated in Fig.~\ref{fig8}.
A comparison of this figure with Fig.~\ref{fig4}
reveals several notable changes, the most dramatic being the
increase in the magnitude of the gaps at both the zone center and 
zone boundary.  
As we have already discussed in Sec.~\ref{weak}, the size of the gap 
is related to the magnitude of the density Fourier coefficients,
so this result is to be expected.  We also notice that the 
$q_x$-dispersion of the lowest branch
is much flatter than for the uniform gas.
The explanation for this behaviour is once again found in
Appendix B where we show that the long-wavelength frequency of plasmons
propagating in the $x$ direction is given by $\omega^2 (q_x ,
q_y = 0) = 2\pi \overline{n}_{2D}(q_x /m_x)$, where $m_x$ is the
effective band mass at the zone center. Since $m_x$ increases
with $V_0$, the dispersion of the plasmon becomes weaker with
increasing modulation.

In Fig.~\ref{fig9} we present the $q_y$-dispersion analogous to
that shown in Fig.~\ref{fig5} for the case of weak modulation.
As expected, the characteristic $\sqrt{q_{y}}$ behaviour is seen
to persist (the ``1-6" branch), but now the 2D plasmon crosses 
the lowest
$q_x = \pi/a$ mode (the ``2-8" branch). At these crossing points, 
the dispersion of the plasmon band as a function of $q_x$
is weak since the frequencies
at $q_x = 0$ and $q_x = \pi/a$ are coincidentally equal. As for
the case of weak modulation, we also see the lowest pair of 
$q_x = \pi/a$ modes (branches ``2-8" and ``3-7") crossing, 
implying that the lowest zone boundary gap
closes at some finite value of $q_y$. This again is possible
due to the different parities of these modes (``2" and ``3") in
the middle panel of Fig.~\ref{fig10}. The same behaviour is now
seen for the next pair of $q_x = \pi/a$ modes which also have
opposite parity. 
The parity of the modes is of course preserved as a function of
$q_y$, but this is not the case as a function of $q_x$. The
mode density ``1" at $q_y = 0$ in Fig.~\ref{fig10}
evolves continuously into the mode density ``2" as $q_x$
is increased from 0 to $\pi/a$, and the parity of the mode
remains unchanged. On the other hand, at $q_y = 3\pi/a$, the
mode density ``6" evolves continuously into ``7", and thus a 
change in parity is observed as the mode disperses with respect
to $q_x$. This change in behaviour is associated with the
crossing of the pair of $q_x = \pi/a$ branches ``2-8" and ``3-7".

We again see in Fig.~\ref{fig9} the merging of zone
boundary and zone center mode frequencies at large $q_y$ values. 
This behaviour is even clearer than in the case of weak
modulation and the explanation is the same.
The mode densities of the lowest two
modes at $q_y = 3\pi/a$ are labeled ``6" and ``7" in 
Fig.~\ref{fig10}; the density profiles of
these two modes have a very similar shape near the cell boundaries
where the equilibrium density is smallest, and only differ 
in the relative sign of the density fluctuation from one cell to the
next. The weak interaction between the density fluctuations
localized at adjacent cell boundaries accounts for the near
degeneracy of these modes. The same can be said of the pair of
density fluctuations labeled ``8" and ``9".

The nature of the mode densities at the lowest zone center gap
is shown in the lower-left panel of Fig.~\ref{fig10}. The mode
labeled ``4" is an odd-parity mode and contributes a peak to the
power absorption near $\omega \simeq 0.25$ a.u.
as shown by the curve labeled ``(b)'' in Fig.~\ref{fig7}. It
is this mode which eventually evolves into the center of mass
mode in the limit of isolated 1D wires. There is an additional
weak peak at $\omega \simeq 0.5$ a.u. corresponding to the
next odd-parity mode at the zone center. It is clear that increasing
the modulation has led to a decrease in the amplitude of the
Drude peak at $\omega = 0$ and a transfer of oscillator strength 
to the other zone center modes.

\subsection{Strong Modulation}
\label{strong}

The system is in the strong modulation regime when the phase
point lies above the solid curve in Fig.~\ref{fig2}.  In particular,
the phase point ``(c)'' corresponds to an equilibrium density
which is made up of an array of isolated quantum wires.
Since there is no appreciable density overlap between adjacent unit 
cells, we expect a relatively flat $q_x$-dispersion for the lower
plasmon bands.  In Fig.~\ref{fig11}, we show the $q_x$-dispersion 
for this case of
strong modulation.  Notice that the lowest plasmon branch
has been pushed down to $\omega \approx 0$ (because of the large
effective band mass $m_x$) and is dispersionless.  
Somewhat surprisingly, the next higher branch exhibits a
dispersion which is larger than the next three branches lying
between $\omega = 0.5$ and $\omega = 0.9$ a.u.
This would not be the expected behaviour in a
tight-binding model of electronic band structure in which the
dispersion of the bands increases with increasing energy (see
Fig.~\ref{fig3}). This
conventional band structure effect is beginning to be evident at
the top of Fig.~\ref{fig11}. As we shall see, the anomalous
dispersion of the first excited band is due to inter-wire
Coulomb interactions. 

The dispersion of the plasmons in the $q_y$-direction is shown
in Fig.~\ref{fig12}. In contrast to the weak and moderately
modulated cases, there are now two gapless modes present in the
spectrum\cite{footnote1}. 
The long-wavelength $\sqrt{q_y}$ character of the $q_x
=0$ 2D plasmon ``1" is still clearly apparent but there is a second
mode ``2" with $q_x = \pi/a$ which has a linear 
dispersion\cite{footnote2}. This
branch originates from the ``2-8" branch in Fig.~\ref{fig9} and in
fact corresponds to a 1D plasmon propagating along each wire,
but with a density fluctuation which changes sign from one
wire to the next. The two lowest curves in Fig.~\ref{fig12}
define the plasmon band for an array of quantum wires and is
qualitatively similar to the results found previously for a
quantum wire superlattice based on the RPA\cite{dassarma85,zhu88}
or hydrodynamic models\cite{eliasson86b}.

The next mode ``4" in Fig.~\ref{fig12} is the 
center of mass (CM) mode for a 1D
wire and the ``4-9" and ``3-7" branches define the CM plasmon
band. This band is qualitatively similar to that found by
Eliasson {\it et al.}, Ref. [\onlinecite{eliasson86b}].
The density fluctuation at $q_x = 0$ and $q_y =0$ is shown
by ``4" in Fig.~\ref{fig13} and, to an excellent approximation,
is simply the derivative of the equilibrium density profile
shown by curve ``(c)" in Fig.~\ref{fig1}. This is the expected
behaviour of the `sloshing' CM mode for parabolic 
confinement\cite{brey89,yip91,dobson94}.
As one disperses along $q_x$ to ``3" at $q_x = \pi/a$, or along
$q_y$ to ``9" at $q_y = 3\pi/a$, one sees that the density
fluctuations at all positions are essentially the same. Thus,
this mode is an intra-wire mode with very little interaction
between the wires.
 
Another interesting feature in Fig.~\ref{fig12} is the way in which
the plasmon band merges with the CM band for large $q_y$. 
This is reminiscent of the mode dispersions
found in 3D parabolic wells \cite{zaremba94} where the 
surface plasmon merges with the CM mode at higher wavevectors.  
The reason for the merging is the same as in the weak and
moderate modulation cases; the density fluctuations shown by
``7" and ``8" in Fig.~\ref{fig13} are symmetric and
antisymmetric versions of each other with very similar density
profiles at the edge of the wires. At this large $q_y$ value, the
density fluctuations on either side of the wire interact with
each other only weakly, as do the fluctuations on different
wires.

We turn next to the explanation of the finite bandwidth of the
``3-4" branch in Fig.~\ref{fig11}. The origin of the dispersion
is the dipole-dipole interaction between the wires. We consider
the limiting case where the wire width $W$ is small compared to
the inter-wire separation $a$. For $q_y = 0$, the electric field
experienced by the $n$-th wire in the dipole approximation is

\begin{equation}
E_n = \frac{2}{a^2} {\sum_s}' \frac{p_{n+s}}{s^2}~,
\label{local_field}
\end{equation}
where $p_n$ is the dipole moment per unit length and the sum
over $s$ excludes the $s=0$ term.
We now suppose that the dipole moment has a plane wave modulation
in the $x$-direction: $p_n = p_0 e^{iq_x na}$. In this case, the
electric field is given by

\begin{equation}
E_n = {2p_0 \over a^2}{\rm e}^{iq_x na} {\sum_s}' \frac{ 
{\rm e}^{iq_x s a}}{s^2}~.
\label{local_field2}
\end{equation}
The quantities $p_n$ and $E_n$ are connected through the
relation $p_n = \alpha (\omega) E_n$ where $\alpha (\omega)$ is
the dipole polarizability of the wire. Since the confining
potential for the wire is roughly parabolic, the polarizability
is given by $\alpha (\omega) = \lambda/(\omega_0^2 - \omega^2)$,
where $\lambda = a\overline{n}_{2D}$ is the line density of the 
wire and $\omega_0 \simeq \sqrt{2\pi^2 V_0/a^2}$
is the frequency of the harmonic potential.
Using this result in Eq.~(\ref{local_field2}), we find the following 
dispersion relation for the CM band:

\begin{equation}
\omega^2 = \omega_0^2 - {4\lambda \over a^2} \sum_{s=1}^{\infty}
\frac{{\rm cos}(q_x a s)}{s^2}~.
\end{equation}
This result is entirely
consistent with the observed $q_x$ dispersion for the ``3-4"
branch shown in Fig.~\ref{fig11}.

Finally, we discuss the FIR power absorption in Fig.~\ref{fig7}.
Each successive curve corresponds to a modulation increase of 
$\Delta V_0 = 1.0$ a.u., starting at $V_0 = 2.0$ a.u. The curves
labeled ``(a)" and ``(b)" were discussed earlier in the context
of weak and moderate modulation, respectively. We note that an
increase in the strength of the modulation transfers oscillator 
strength from the Drude peak to the higher $q_x = 0$ modes
having a nonzero dipole moment. The two peaks which grow in
intensity between curves ``(a)" and ``(b)" evolve from the
$q=2\pi/a$ and $q=4\pi/a$ 2D bulk plasmons. Both peaks are
red-shifted with increasing modulation with respect to their
positions in the uniform gas limit. Near the curve labeled
``(b)", this trend is reversed and the peak frequencies begin to
increase with further increase in modulation. At the same
time, the oscillator strength of the lower peak continues to
increase, while that of the higher frequency peak begins to
decrease. Beyond this point we also see a rapid reduction in the
strength of the Drude peak at $\omega = 0$.

The change in behaviour of the peak positions and intensities
near the curve ``(b)" occurs as the 2D-1D phase boundary in 
Fig.~\ref{fig2} is being crossed. Thus, a signature of the 2D
$\to$ 1D transition is the observation of a minimum in the
fundamental resonance frequency as a function of modulation, and
the disappearance of the Drude peak. Beyond this minimum, the
fundamental resonance grows in intensity as it evolves into the
CM (Kohn) mode\cite{brey89,yip91,dobson94}
for parabolic confinement. For large modulations,
the quantum wires are well-separated and
sit in an effectively harmonic potential with
frequency $\omega_0 \simeq \sqrt{2\pi^2 V_0/a^2}$, which
explains the increasing trend observed in Fig.~\ref{fig7}. This
also explains why the fundamental resonance eventually exhausts
the total FIR oscillator strength.

As an overview of our results, we plot in Fig.~\ref{fig14} the
zone boundary and zone center frequencies at $q_y = 0$ as a function of the
strength of the modulating potential. Only the lowest ZB modes
exhibit a linear separation of the frequency with $V_0$ and is
in accord with our perturbation theory results. The lower branch
is seen to tend to zero as $V_0$ increases as this mode evolves
into the ZB (or out-of-phase) plasmon of a 1D wire superlattice.
The lowest ZC mode (indicated by solid circles) tracks along the
CM mode in Fig.~\ref{fig7} and, as discussed above, goes through
a minimum at the transition from 2D to 1D behaviour. This
behaviour is qualitatively similar to that 
observed\cite{hansen87,brinkop88}, although it should be noted
that in the experiments the average density of the gas
decreased as a function of the gating potential, whereas our
results are for a constant density. The CM mode crosses the
upper branch of the second ZB mode at $V_0/E_f \simeq 20$ and
the pair then defines the CM band with its dipole induced
dispersion. For the higher modes we see the merging of 
successive pairs of ZC 
and ZB modes in the 1D limit. The increasing frequency of
these modes is once again due to the increasing curvature of the
confining potential. Fig.~\ref{fig14} extends the variation of
the mode frequencies with $V_0$ into the localized 1D regime
which is beyond that given previously\cite{eliasson86a,cataudella88}.
In particular, we do not find the unphysical result claimed in
Ref. [\onlinecite{cataudella88}] that the system does not
support plasmons in the strong localization limit.

\section{Conclusions}
\label{conclusions}

In this paper, we have demonstrated that the Thomas-Fermi-Dirac-von
Weizs\"acker approximation provides a realistic description of the
collective excitations in a modulated 2DEG.
We have presented a detailed investigation of the equilibrium
properties, and have numerically mapped out
the parameter space that defines the transition from
2D to 1D behavior.  
We have also calculated the plasmon dispersions for propagation 
along and across the modulation direction in the weak, moderate
and strong modulation limits.  In agreement with earlier work
on this problem, the modulation of the equilibrium
density leads to the appearance of gaps in the plasmon dispersion
in the direction of the modulation;
as the modulation becomes stronger the bands become narrower and
the gaps larger.  For propagation in a direction perpendicular
to the direction of modulation, we have shown that the
long-wavelength 2D plasmon is unaffected by the modulation
potential. However, at shorter wavelengths, the plasmon
dispersions exhibit interesting behaviour as a function of
$q_y$ and the
explicit calculation of the mode densities provides a more
complete understanding of the physical nature of these excitations.

The power absorption in the long-wavelength limit has also been
calculated for a range of modulation potentials.  We have found that
the oscillator strength is predominantly in the Drude peak for
weak modulation and shifts to higher dipole modes as the
modulation is increased. The reduction of the Drude peak
is one useful indicator to gauge the effective dimensionality of 
the system. We have also shown that the 2D bulk plasmon at $q =
2\pi/a$ evolves continuously with increasing modulation into the
CM mode of a quantum wire superlattice. The frequency of the
mode at first decreases, and then passes through a minimum as
the 2DEG makes the transition from a continuous charge
distribution to an array of isolated wires. This minimum is a
second signature of the 2D $\rightarrow$ 1D  crossover.
At large modulations, the confining potential is effectively
harmonic, and the FIR oscillator strength resides in the CM mode
in accord with the generalized Kohn theorem.

We are at present completing a detailed analysis of the magnetoplasma
excitations in laterally modulated 2D systems.  These results
will be reported elsewhere.

\acknowledgments
 
The work was supported by a grant from the Natural Sciences and
Engineering Research Council of Canada. We would like to thank Dr. D. A. 
W. Hutchinson for useful discussions.

\appendix
\section{Calculation of Plasmon Band Gaps}
\label{perturbation}

In this appendix, we determine the eigenvalue spectrum
of a laterally modulated 2DEG in the weak modulation regime.  
The eigenvalue problem that we wish to solve is 
\begin{eqnarray}
\omega^2 c_{G} &=& \sum_{G'G''} B_{GG'} \tilde{M}_{G'G''} 
c_{G''}\nonumber \\ &=& \sum_{G'} N_{GG'} c_{G'}~,
\label{app_matrix}
\end{eqnarray}
with
\begin{eqnarray}
N_{GG'} &\equiv& \sum_{G''} B_{GG''}\tilde{M}_{G''G'}~.
\label{N_matrix}
\end{eqnarray}
In the absence of an external modulation, the $N$-matrix is 
diagonal. Denoting the matrices in this limit by a superscript 0,
we have
\begin{eqnarray}
B^0_{GG'} &=& \frac{1}{2}((q_x + G)^2 + q_y^2)\delta_{GG'} \nonumber \\
\tilde{M}^{0}_{GG'} &=& \left [ \frac{4 \pi
\overline{n}_{2D}}{\sqrt{(q_x + G)^2 + q_y^2}} + 4 C_1
\overline{n}_{2D} - \frac{3}{2} C_3 \sqrt{\overline{n}_{2D}} +
\frac{\lambda_w}{2}\left ((q_x + G)^2+q_y^2\right  )\right 
]\delta_{GG'}~.
\label{B0M0}
\end{eqnarray}
With these results, we have at $q_y = 0$,
\begin{eqnarray}
N^0_{GG'} &=& \sum_{G''} B^0_{GG''}\tilde{M}^0_{G''G'} \nonumber \\
&=& \left [2\pi\overline{n}_{2D} |q_x + G| + \left ( 2C_1 
\overline{n}_{2D} - \frac{3}{4}C_3 \sqrt{\overline{n}_{2D}}
\right ) (q_x+G)^2 + \frac{\lambda_w}{4}
(q_x+G)^4 \right ]\delta_{GG'}\nonumber \\
&\equiv& \omega_0^2(q_x+G)\delta_{GG'}~.
\label{diagonal_N}
\end{eqnarray}
This equation gives the plasmon frequency in Sec.~\ref{uniform}.
With $G$ taking on all possible values, $q_x$ can be restricted
to the first Brillouin zone,
$-\pi/a \le q_x \le \pi/a$. 

The introduction of an external modulation lifts the degeneracies 
in the uniform gas spectrum at
the zone center ($q_x =0$) and at the zone boundary ($q_x =\pi/a$).
At the zone center (ZC), the reciprocal lattice vectors
coupling degenerate modes are given by
\begin{equation}
q_x = 0,~~ G_1 = n\frac{\pi}{a},~~G_2=-n\frac{\pi}{a}~,
\label{zc}
\end{equation}
where $n$ is an even integer, while at the zone boundary (ZB)
\begin{equation}
q_x=\frac{\pi}{a},~~G_1=(n-1)\frac{\pi}{a},
~~G_2=-(n+1)\frac{\pi}{a}~,
\label{zb}
\end{equation}
where $n$ is an odd integer.  At either 
ZC or ZB, $|q_x+G_1|=|q_x+G_2|=n{\pi}/{a}\equiv q_n$ and 
$(G_1-G_2) = 2n{\pi}/{a} =2q_n$.

We now apply degenerate perturbation theory 
to obtain first--order corrections to the frequencies at these
points in the first Brillouin zone. Retaining only the two
degenerate modes of interest, we have the pair of
equations
\begin{eqnarray}
\omega^2 c_{G_1} &=& N^0_{G_1G_1}  c_{G_1} + N^1_{G_1G_2} c_{G_2} \nonumber \\
\omega^2 c_{G_2} &=& N^1_{G_2G_1}  c_{G_1} + N^0_{G_2G_2} c_{G_2} ~,
\label{red_eigenvalue}
\end{eqnarray}
where $ N^1_{G_1G_2} = N^1_{G_2G_1}$ is the lowest order
off-diagonal correction to the $N$-matrix. It is given by
\begin{equation}
N^1_{G_1G_2} =
B^0_{G_1G_1}\tilde{M}^1_{G_1G_2} + B^1_{G_1G_2}\tilde{M}^0_{G_2G_2}~,
\label{red_Nmatrix}
\end{equation}
where
\begin{eqnarray}
\tilde{M}^1_{G_1G_2} &=& \frac{8\pi}{q_n}\sqrt
{\frac{\overline{n}_{2D}}{a}} \overline{\psi}_0[2q_n]
+ \frac{4C_1}{\sqrt{a}} \overline{n}_0[2q_n] -
\frac{3C_3}{\sqrt{a}} \overline{\psi}_0[2q_n] -
\frac{2\lambda_w}{\sqrt{\overline{n}_{2D}a}}q_n^2\overline{\psi}_0[2q_n]
\nonumber \\ B^1_{G_1G_2} &=& -\frac{2}{\sqrt{\overline{n}_{2D}}}q_n^2 
\overline{\psi}_0[2q_n] ~.
\end{eqnarray}
In the above, $\overline{\psi}_0[2q_n]$ and $\overline{n}_0[2q_n]$
denote the $n$-th Fourier components of the groundstate
wavefunction and charge density respectively.
Replacing $\overline{\psi}_0[2q_n]$ by 
$\overline{n}_0[2q_n]/2\sqrt{\overline{n}_{2D}}$,
Eq.~(\ref{red_Nmatrix}) can be cast in the form
\begin{eqnarray}
N^1_{G_1G_2} = \left [ -\omega_0^2(q_n) + 
\frac{3}{8}C_3\sqrt{\overline{n}_{2D}}
q_n^2  - \frac{3}{4}\lambda_w q_n^4 \right ] \left [ 
\frac{\overline{n}_0[2q_n]}{\overline{n}_0[2q_0]} \right ]~.
\end{eqnarray}
This implies that the eigenvalues of (\ref{red_eigenvalue}) are given by
\begin{eqnarray}
\omega_{\pm}^2 (q_n) &=& N^0[q_n] \pm |N^1[q_n]| \nonumber \\
&=& \omega_0^2(q_n) \pm \left | \omega_0^2(q_n) - 
 \frac{3}{8}C_3\sqrt{\overline{n}_{2D}}q_n^2  + 
\frac{3}{4}\lambda_w q_n^4 \right | 
\left |\frac{\overline{n}_0[2q_n]}{\overline{n}_0[2q_0]} \right |~,
\label{freq}
\end{eqnarray}
where $N^0[q_n] \equiv N^0_{G_1G_1}$ and $N^1[q_n] \equiv
N^1_{G_1G_2}$.
The size of the gap at the ZC or ZB is given by $2|N_1[q_n]|$.
We have tested the validity of (\ref{freq}), and in the regime 
where the perturbative results are valid, we find excellent 
agreement between the analytical and numerical calculations
as illustrated in Fig.~\ref{fig14}.
In the Hartree approximation, we obtain the simple relation

\begin{equation}
\omega_{\pm}^{2}(q_n) = \omega_{0}^{2}(q_n) \left ( 1 \pm 
\left |\frac{\overline{n}_{0}[2q_n]}{\overline{n}_{0}[2q_0]} \right | 
\right )~.
\end{equation}
This result is entirely consistent with the one obtained by
Krasheninnikov and Chaplik\cite{krasheninnikov81} 
in the long-wavelength limit.

\section{Determination of Long Wavelength Dispersions}
\label{q_dispersions}

A persistent feature of the dispersion relations we have
calculated (see Figs.~\ref{fig4}, \ref{fig8} and \ref{fig11}), is that 
the long--wavelength limit of the
lowest plasmon has the characteristic $\sqrt{q}$ dependence of the
2D-bulk plasmon, regardless of the modulation.
In this section, we show analytically that the long-wavelength
behaviour of the lowest plasmon branch is $\sqrt{q}$ for {\em
any} modulation amplitude.

Once again, the eigenvalue problem that we are solving has the
form

\begin{eqnarray}
\omega^2 c_{G} &=& \sum_{G'G''} B_{GG'} \tilde{M}_{G'G''} c_{G''}\nonumber \\
&=& \frac{1}{\lambda_w}\sum_{G'G''} h_{GG'} \tilde{M}_{G'G''} c_{G''}~,
\label{matrix}
\end{eqnarray}
where $h_{GG'}$ is the Hamiltonian 
which, for $q_y = 0$, determines the groundstate von Weizs\"acker 
wave function $\overline{\psi}_0[G]$.
In the long-wavelength limit, the dominant contribution to the
$\tilde{M}$--matrix comes from the $G''=0$ component of the Hartree 
matrix in Eq.~(\ref{hartree}),
\begin{equation}
\lim_{q\rightarrow 0}\tilde{M}_{GG'} \simeq 
\lim_{q\rightarrow 0} M^{H}_{GG'} \simeq \frac{4\pi}{qa}
\overline{\psi}_{0}[G]\overline{\psi}_{0}[G']\,.
\label{app_hartree}
\end{equation}
With this result, and noting that $\sum_{G'} h_{GG'}
\overline{\psi}_0[G'] = \varepsilon_0({\bf q}) \overline{\psi}_0[G]$,
Eq.~(\ref{matrix}) can be written as
\begin{equation}
\lambda_w \omega^2 c_G =
\frac{4\pi}{qa} \varepsilon_{0}({\bf q}) \overline{\psi}_0[G]
\sum_{G''}\overline{\psi}_0[G'']c_{G''}\,.
\label{matrix2}
\end{equation}
The eigenvalues $\omega^2$ of (\ref{matrix2}) 
are thus given by

\begin{eqnarray}
\omega^2 (q_x,q_y) = \frac{4\pi\overline{n}_{2D}}{q}
\varepsilon_{0}({\bf q})\,,
\end{eqnarray}
since $\overline n_{2D} = a^{-1} \sum_G|\overline{\psi}_0[G]|^2$.
For small ${\bf q}$ we have $\varepsilon_{0}({\bf q}) \simeq
{\textstyle{\lambda_w}\over \textstyle{2}}\left (
{\textstyle{q_x^2}\over \textstyle{m_x}} + q_y^2\right )$, where
$m_x$ is the effective band mass due to the confining potential
in the modulation direction.
Of interest here is the determination of the dispersions in
both the $q_x=0,~q_y \rightarrow 0$ and $q_y=0,~q_x \rightarrow 0$
limits.  In these two cases we have
\begin{eqnarray}
\omega^2 (q_x=0,q_y) &\simeq& {2\pi\overline{n}_{2D}} q_y 
\label{q_y}
\end{eqnarray}
\begin{eqnarray}
\omega^2 (q_x,q_y=0) &\simeq& 
{2\pi\overline{n}_{2D}\over m_x} q_x~.
\label{q_x}
\end{eqnarray}
Eq.~(\ref{q_y}) illustrates the fact that for any modulation, the 
long--wavelength behaviour of the 2D plasmon is $\sqrt{q_y}$.  
Eq.~(\ref{q_x}) on the other hand
indicates that the $\sqrt{q_x}$ dispersion is suppressed 
with increasing modulation as a result of the 
effective mass $m_x$.

\begin{figure}
\caption{Equilibrium density profiles of Eq.~(\ref{SE}) with 
$v_{\rm ext}$
given by Eq.~(\ref{modpot}).  The curves labeled by (a), (b), (c),
and (d) correspond to $V_{0}/E_{f} = 6.4$, 19.1, 28.6 and 41.4,
respectively. The mean density in all cases is
${\overline{n}_{2D}} = 0.1$.} 
\label{fig1}
\end{figure}

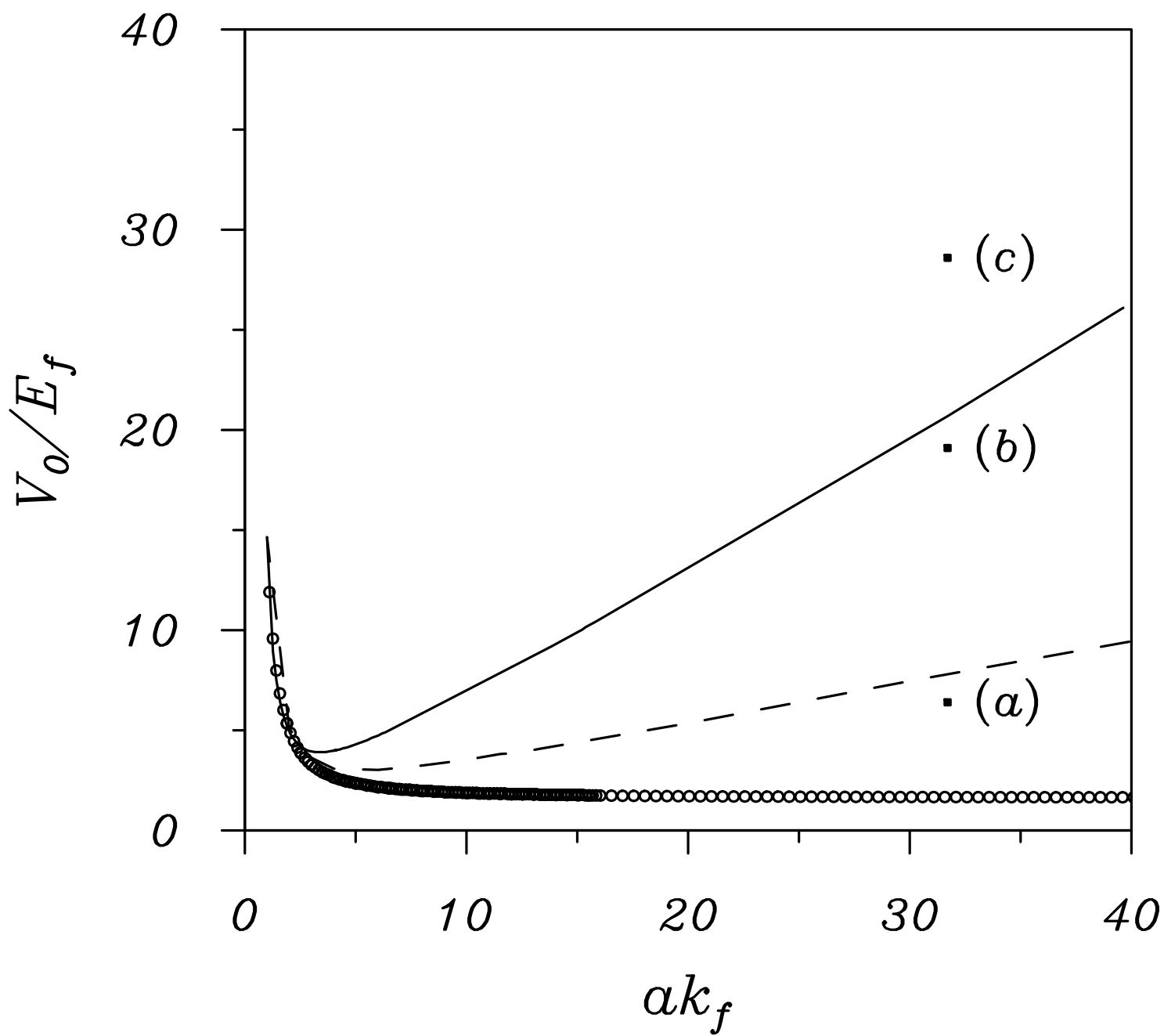
\begin{figure}
\caption{``Phase''-diagram for the 2D $\rightarrow$ 1D transition.
The solid and dashed curves include all interactions with
${\overline{n}_{2D}} = 0.1$ and ${\overline{n}_{2D}} = 1.0$
respectively.  The open circles correspond to the noninteracting phase
curve ({\em i.e.} a confined noninteracting Fermi gas) which is independent of the
average 2D density.}
\label{fig2}
\end{figure}

\begin{figure}
\caption{The calculated TFDW energy bands along the direction 
of modulation.  This figure is evaluated for a moderate modulation corresponding
to ``(b)'' in Fig.~\ref{fig1}.}
\label{fig3}
\end{figure}

\begin{figure}
\caption{The dispersion of mode frequencies with wave vector $q_{x}$
for the weakly modulated 2DEG.  The solid curve is for the uniform 2DEG.
A full explanation of the encircled numbers is given in the
text.}
\label{fig4}
\end{figure}

\begin{figure}
\caption{The dispersion of mode frequencies with wave vector $q_{y}$
for the weakly modulated 2DEG.  Here $q_{x}=0$ (solid circles),
and $q_{x}=\pi/a$ (open circles).
A full explanation of the encircled numbers is given in the
text.}
\label{fig5}
\end{figure}

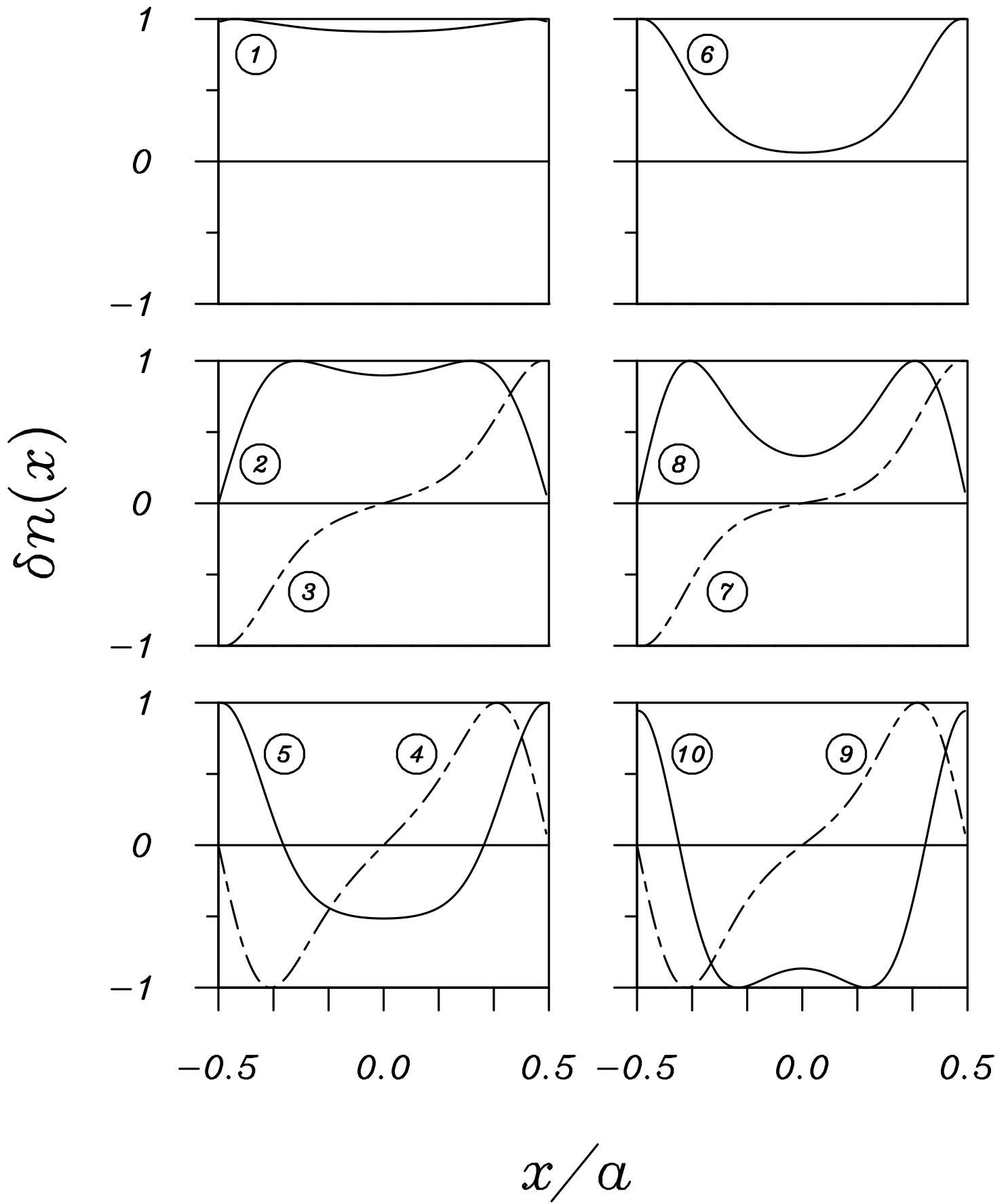
\begin{figure}
\caption{The mode densities in the weak modulation regime for 
various bands.  The left panels are evaluated at fixed $q_y = 0$ and the
right panels at evaluated at $q_y = 3\pi/a$.  The sequence for the 
panels from top to bottom is $q_x = 0$, $q_x = \pi/a$, and $q_x = 0$.  
See the text for full details.}
\label{fig6}
\end{figure}

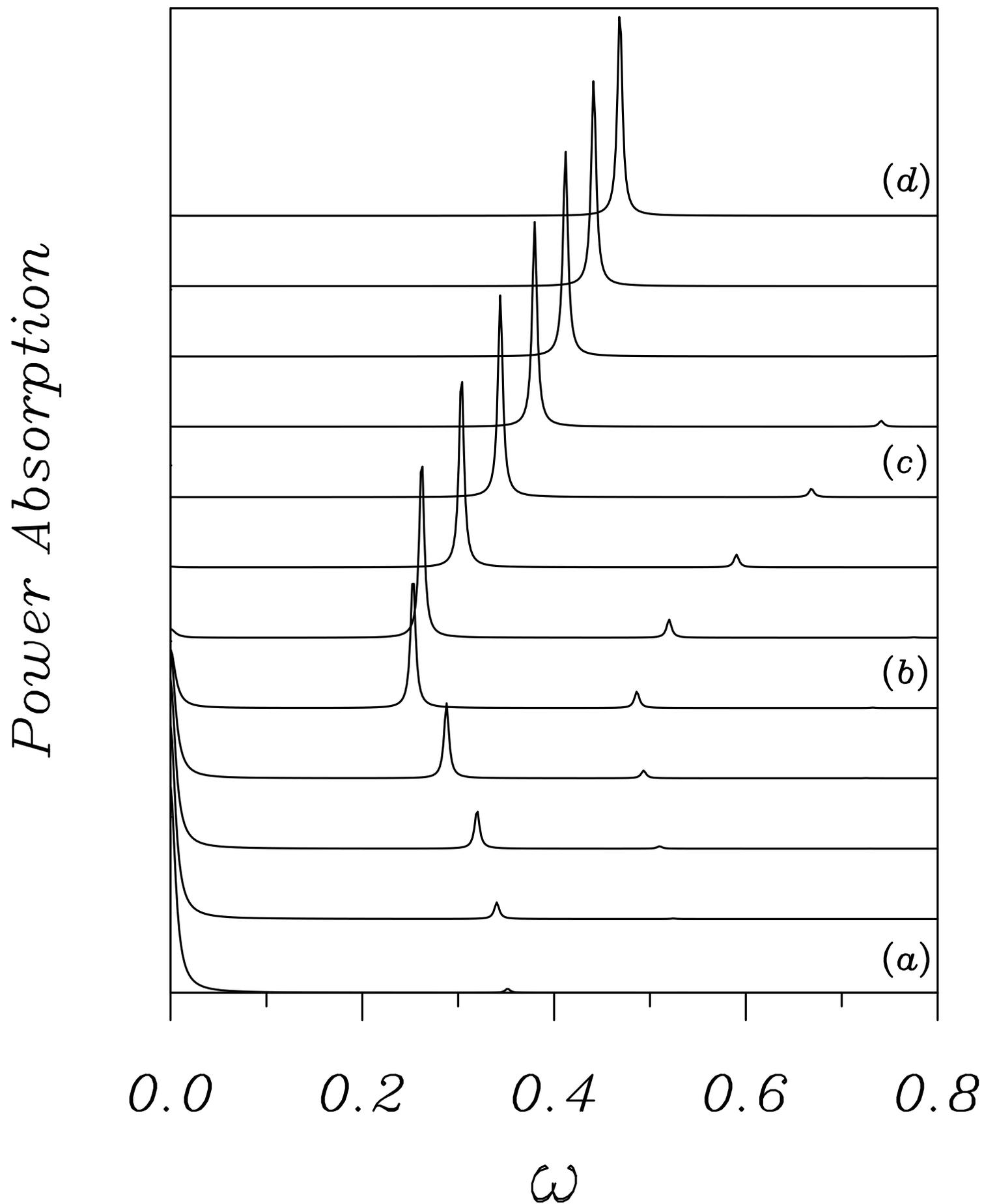
\begin{figure}
\caption{Power absorption as a function of frequency.  The curves
are ordered from weakest (bottom) to strongest (top) modulation and
are offset by $\Delta V_0 = 1.0$ a.u.
The labels $(a),~(b),~(c)$ and $(d)$ are consistent with the
notation adopted in Figs.~\ref{fig1} and \ref{fig2}.}
\label{fig7}
\end{figure}

\begin{figure}
\caption{As in Fig.~\ref{fig4}, but for moderate modulation.}
\label{fig8}
\end{figure}

\begin{figure}
\caption{As in Fig.~\ref{fig5}, but for moderate modulation.}
\label{fig9}
\end{figure}

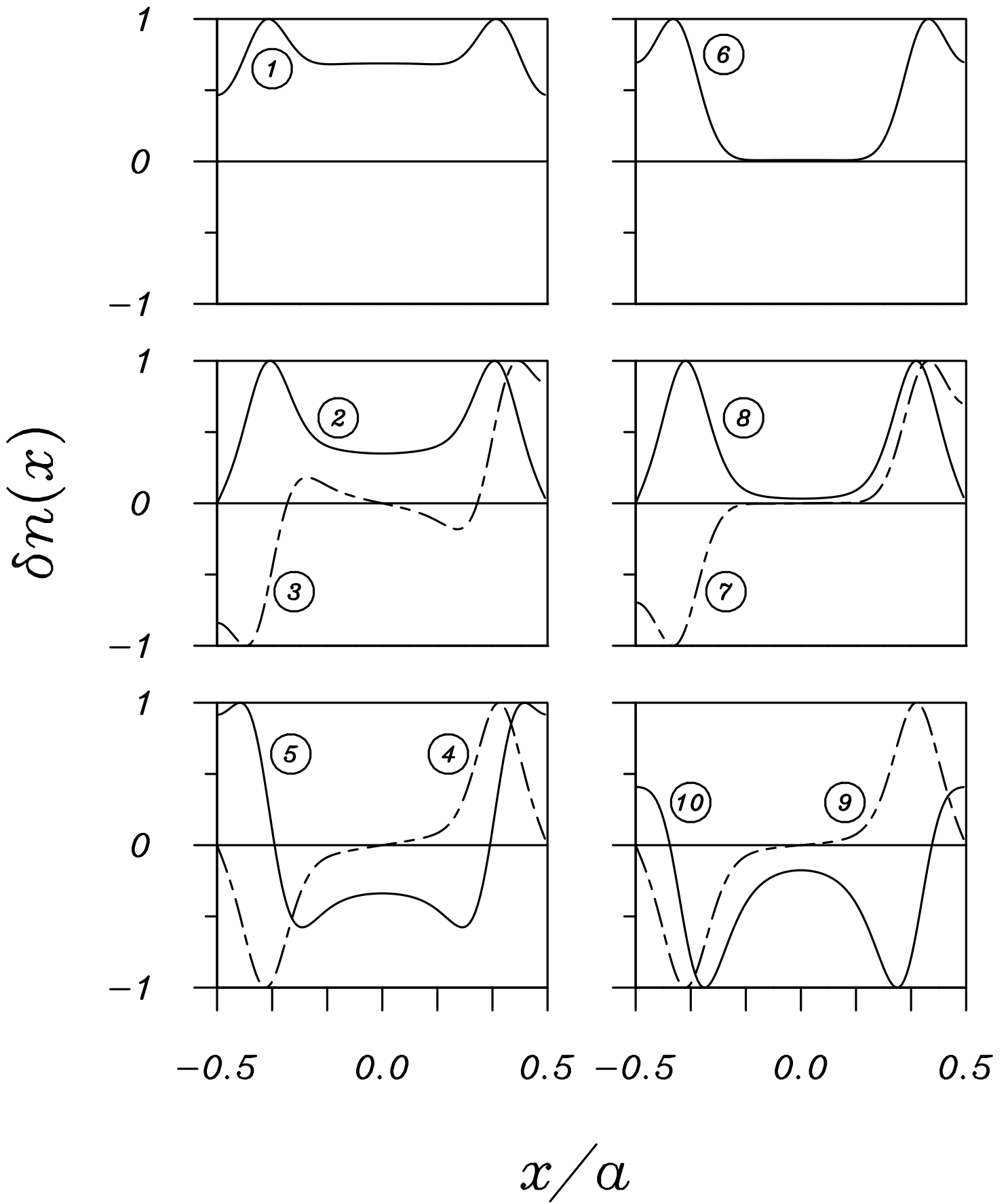
\begin{figure}
\caption{As in Fig.~\ref{fig6}, but for moderate modulation.}
\label{fig10}
\end{figure}

\begin{figure}
\caption{As in Fig.~\ref{fig4}, but for strong modulation.}
\label{fig11}
\end{figure}

\begin{figure}
\caption{As in Fig.~\ref{fig5}, but for strong modulation.}
\label{fig12}
\end{figure}

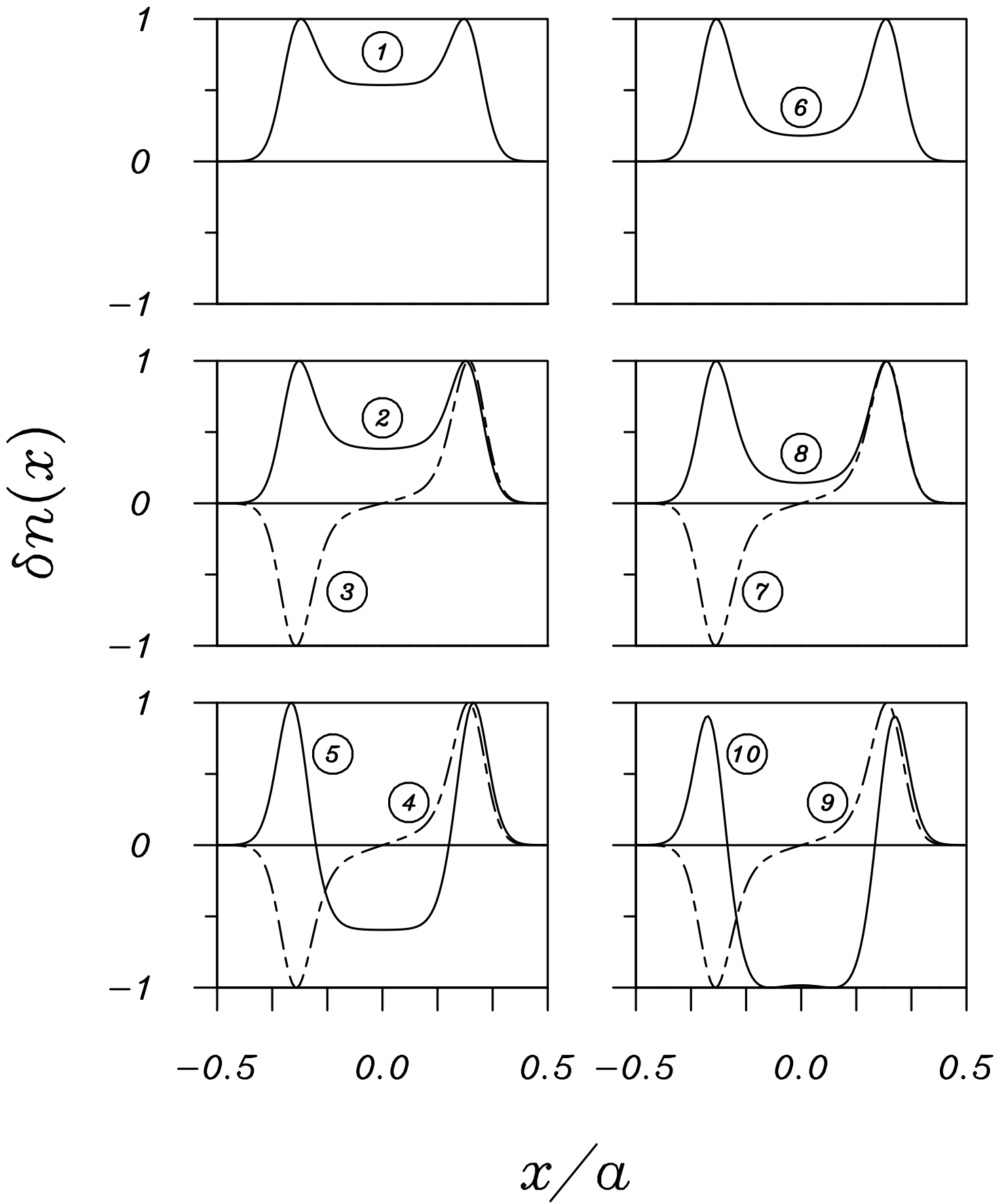
\begin{figure}
\caption{As in Fig.~\ref{fig6}, but for strong modulation.}
\label{fig13}
\end{figure}

\begin{figure}
\caption{Zone boundary (open circles) and zone center (filled circles)
frequencies (in a.u.) at $q_y=0$ as a function of the strength of the modulating
potential. The solid curves are the solutions given by Eq.~(\ref{freq})
for $n = 1$.}
\label{fig14}
\end{figure}

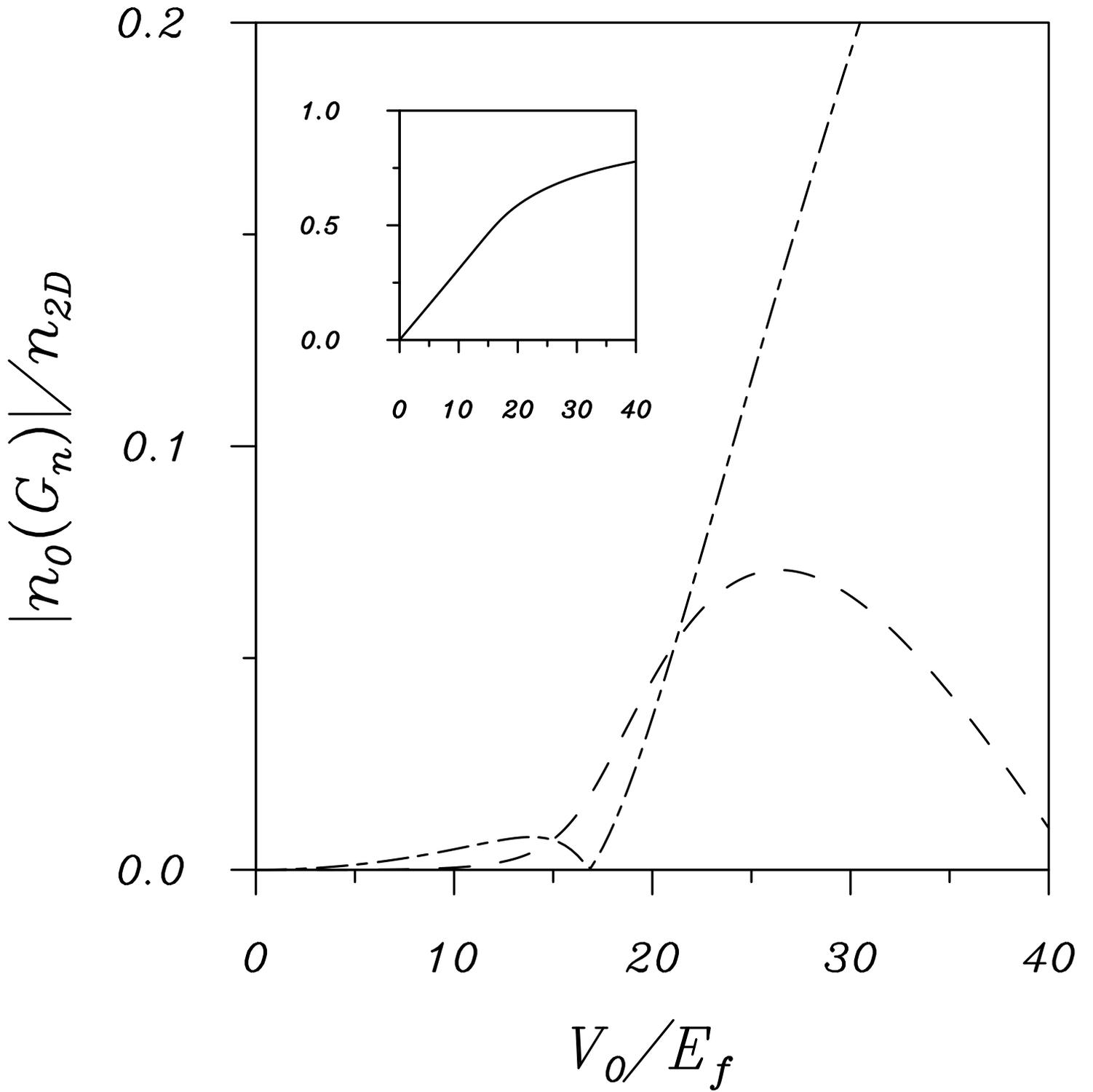
\begin{figure}
\caption{The first three Fourier coefficients of the equilibrium
charge density as a function of modulation strength. The curves are
normalized with respect to the average areal density 
$\overline{n}_{2D}$. Here $G_n = 2n\pi/a$, with $n=1$ corresponding 
to the figure inset (solid line), $n=2$ the short-dash-long-dash 
curve and $n=3$ the dashed curve.}
\label{fig15}
\end{figure}


\begin{references}
\bibitem{footnote1}
The mode is actually gapless at $q_y =0$ for all $q_x$ values
(see the ``1-2" branch in Fig.~\ref{fig11}). Fig.~\ref{fig12}
shows the $q_y$-dispersion for the limiting values of $q_x = 0$
and $q_x = \pi/a$.
\bibitem{footnote2}
For a single wire of width $a$, the plasmon frequency behaves
as\cite{dassarma85,eliasson86b,zhu88} $\omega \propto
qa|\ln(qa)|^{1/2}$. For a pair of such wires, the out-of-phase
plasmon has a linear dispersion\cite{zhu88}. The $q_x = \pi/a$
mode should behave in a similar way.
\end{references}
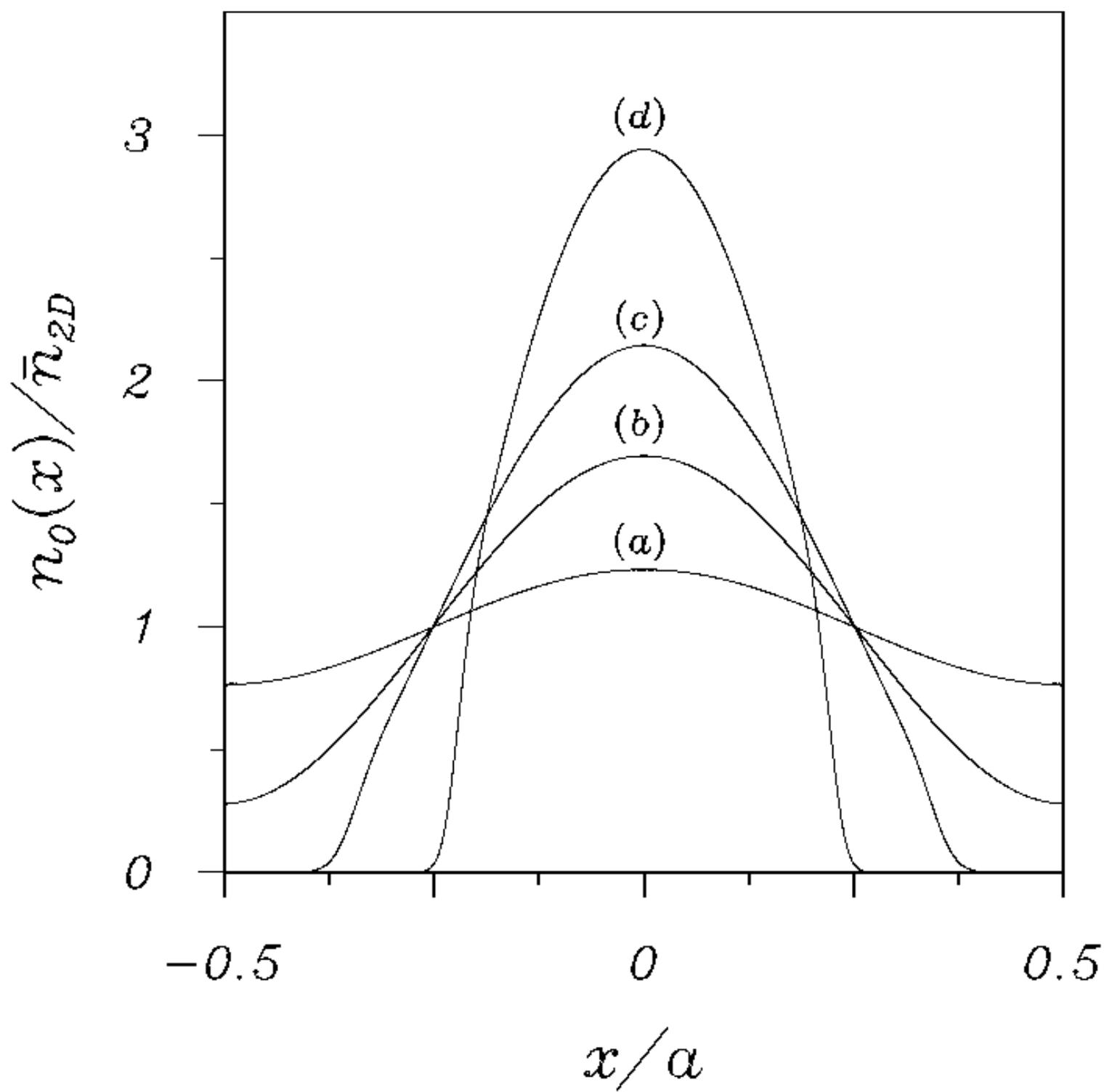
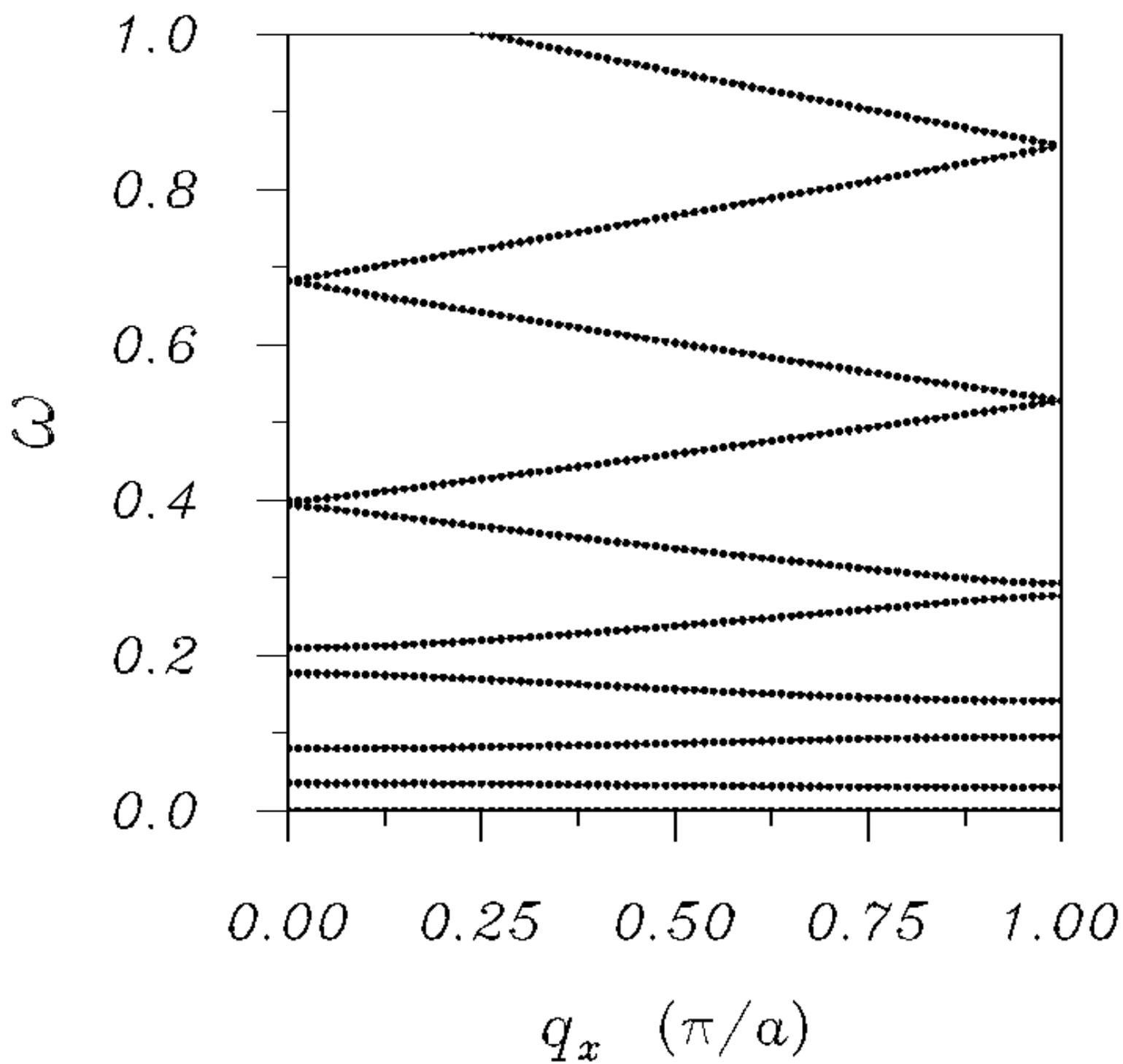
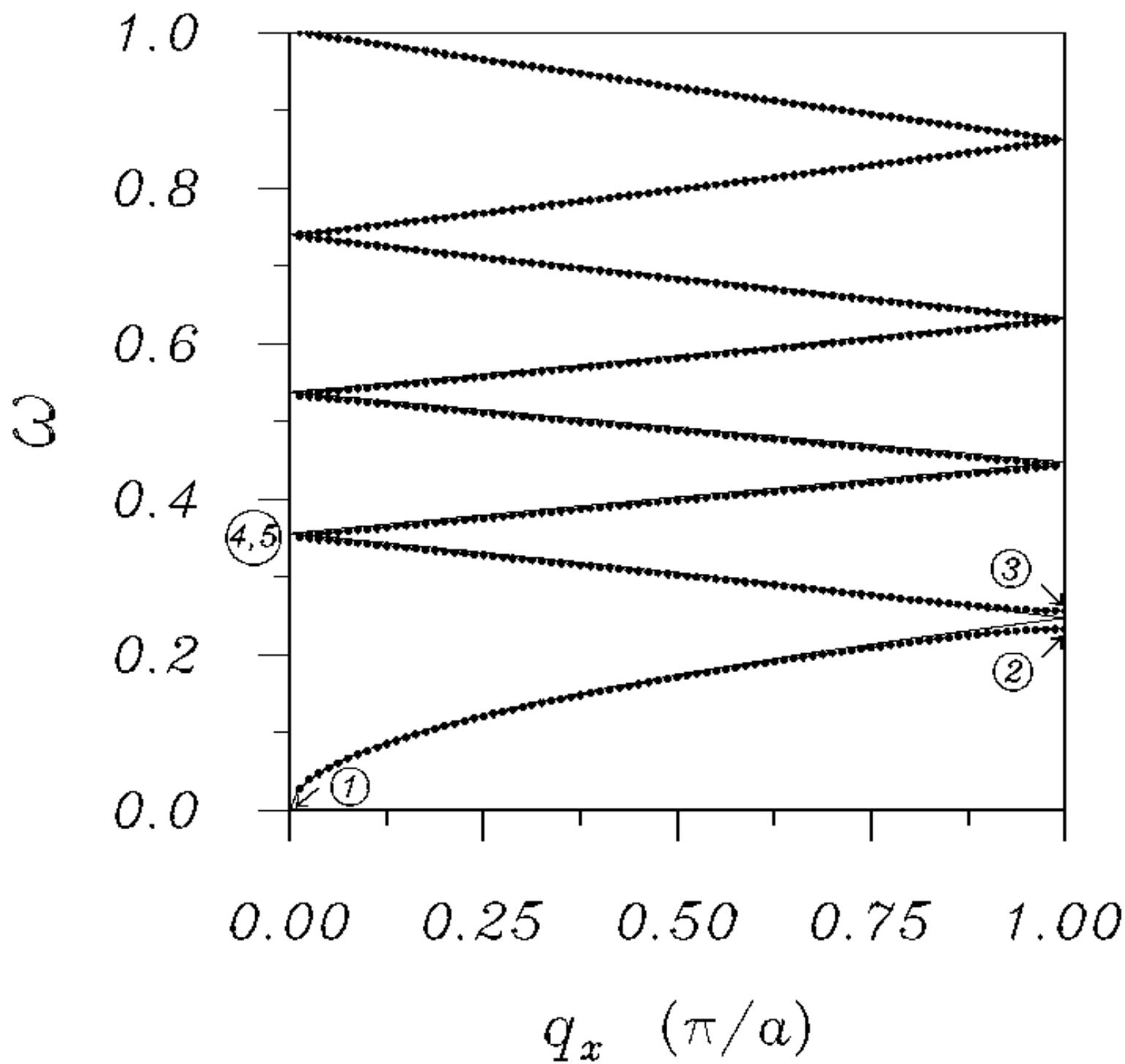
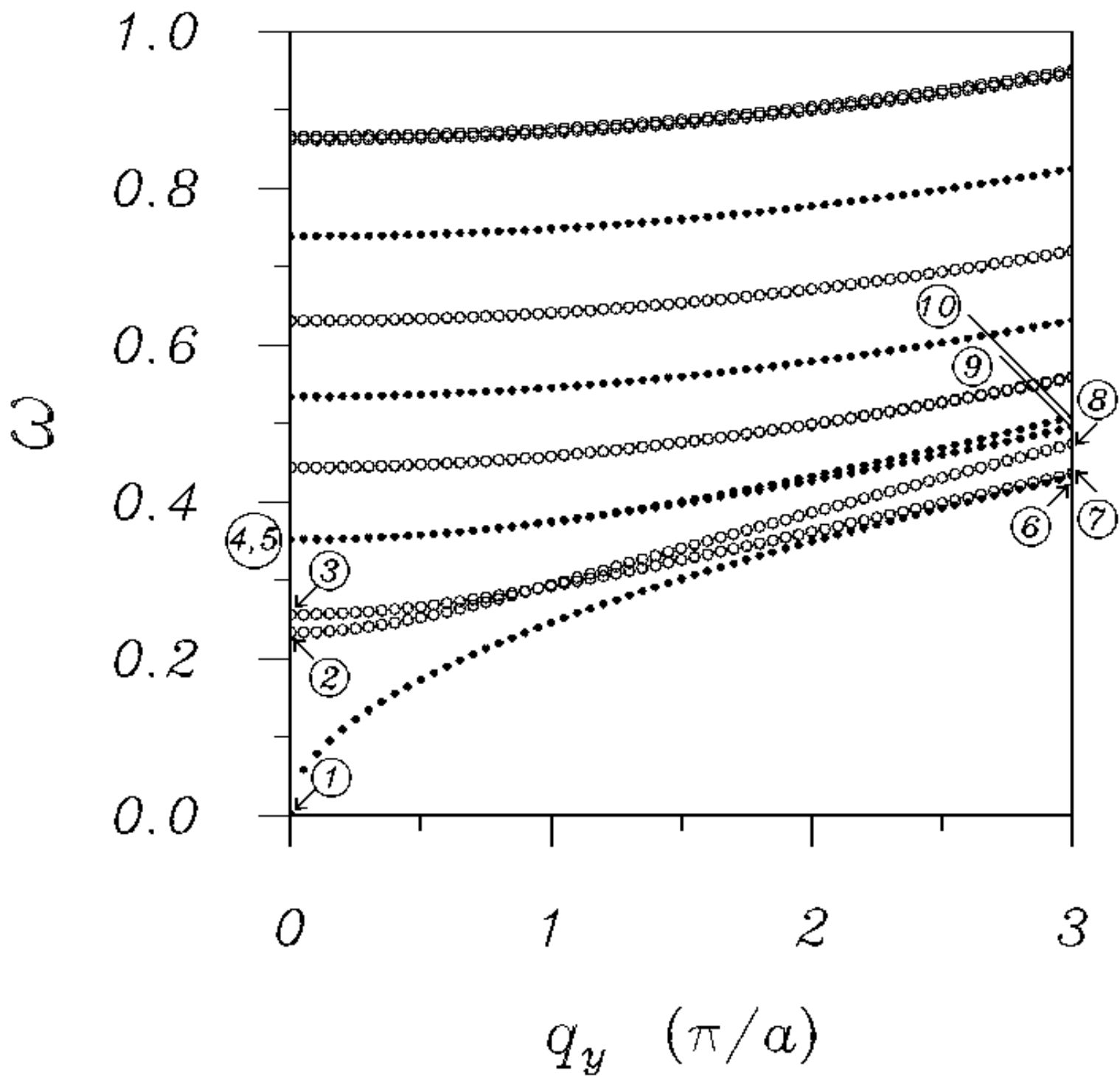
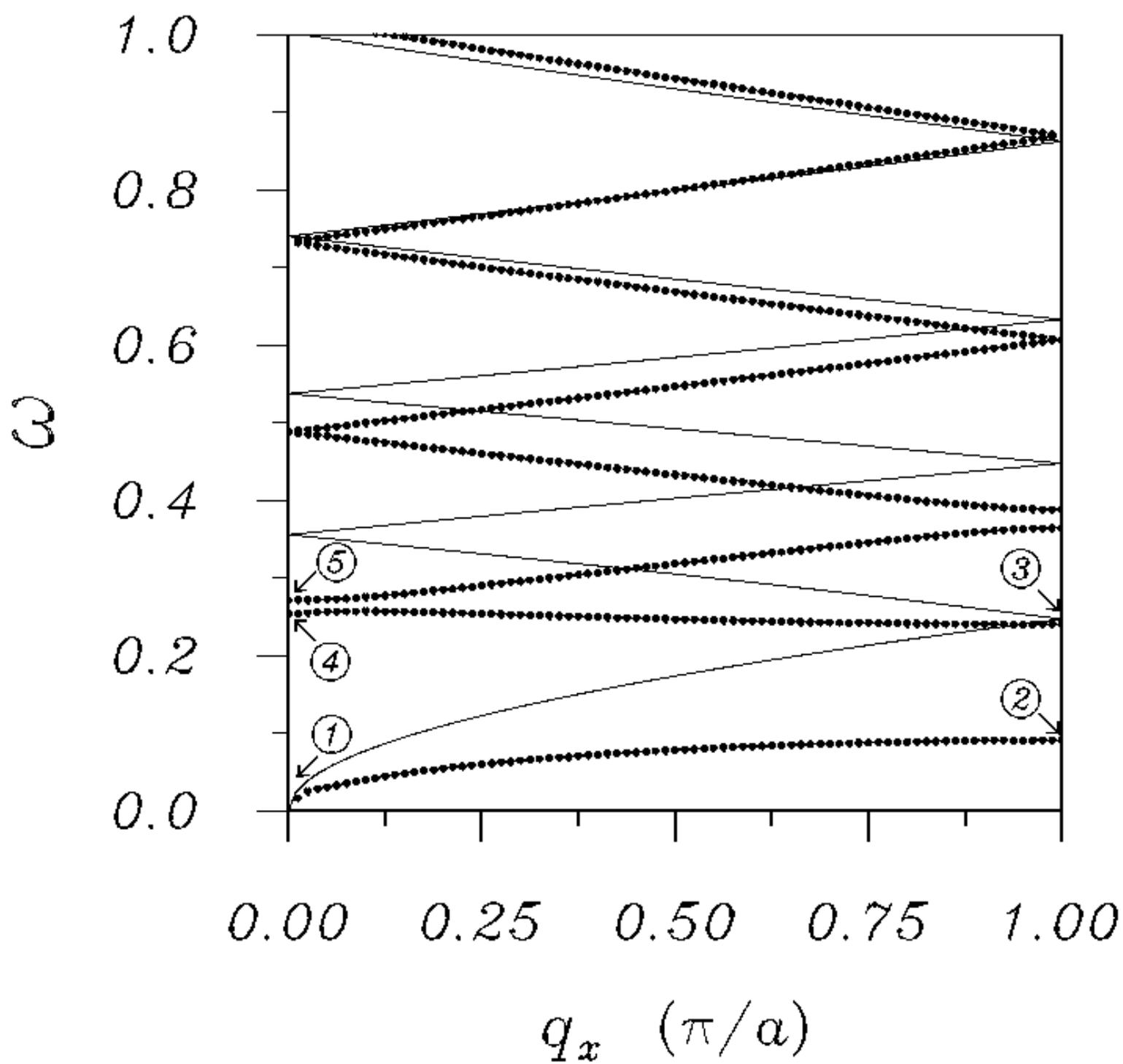
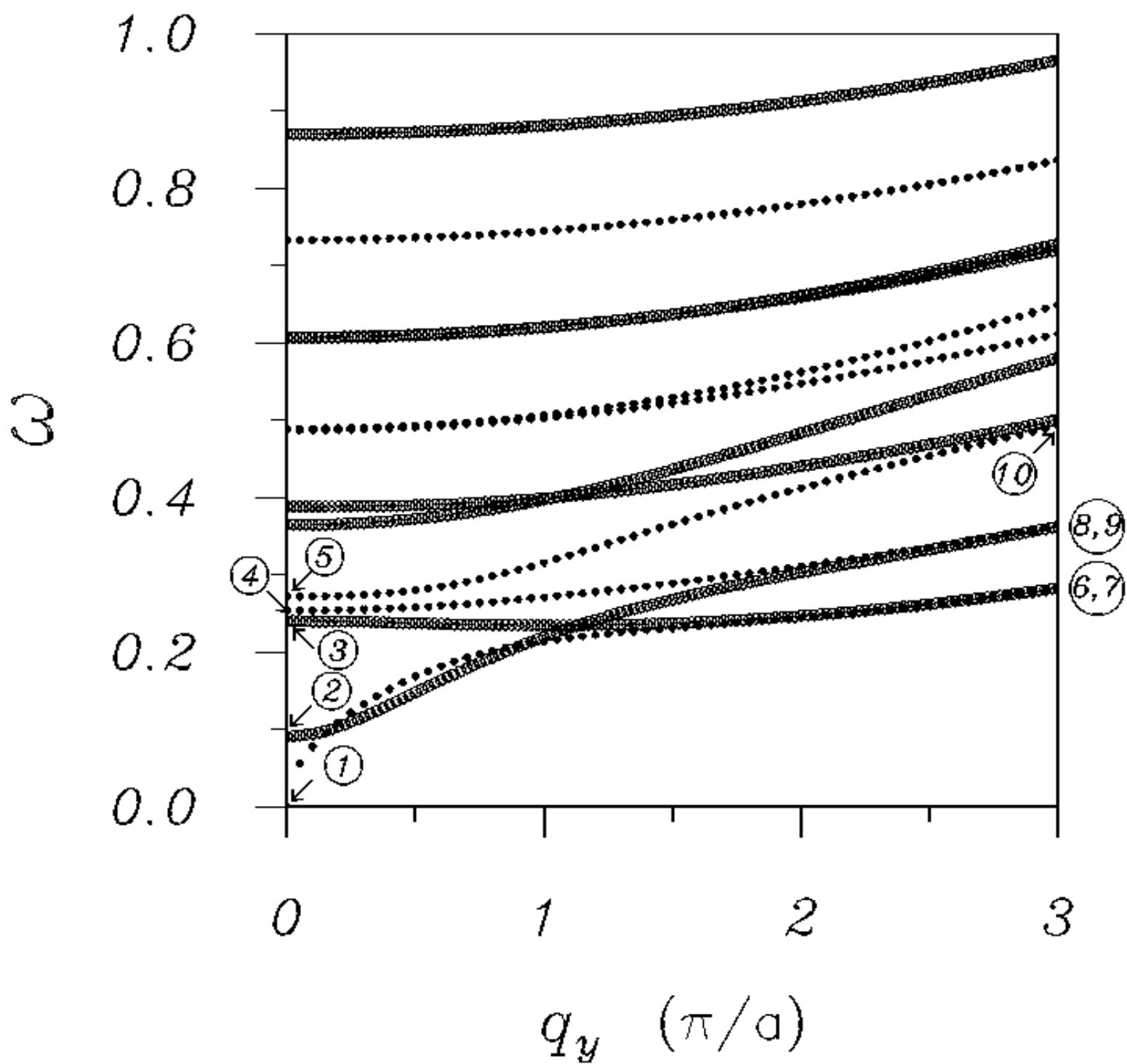
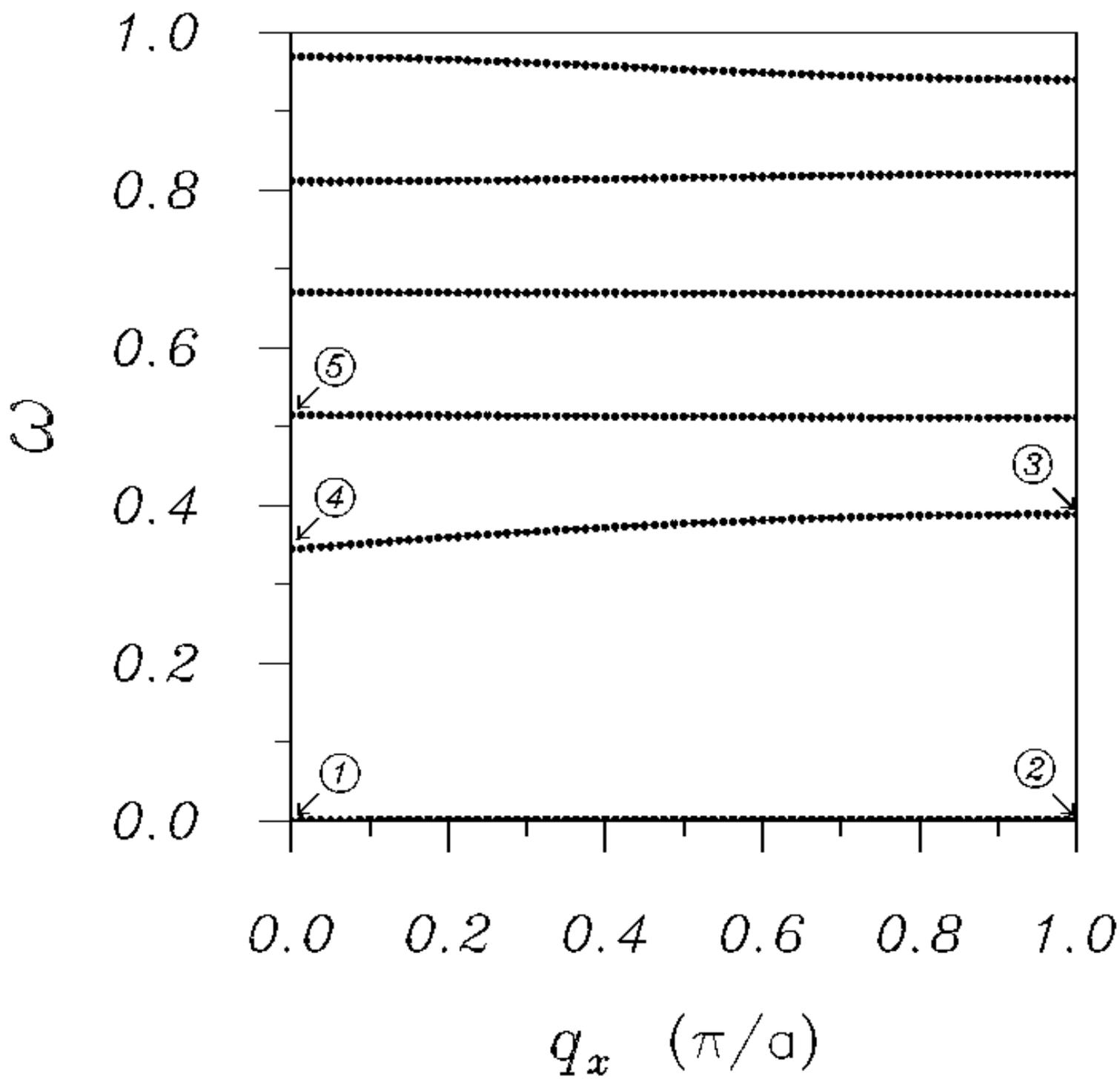
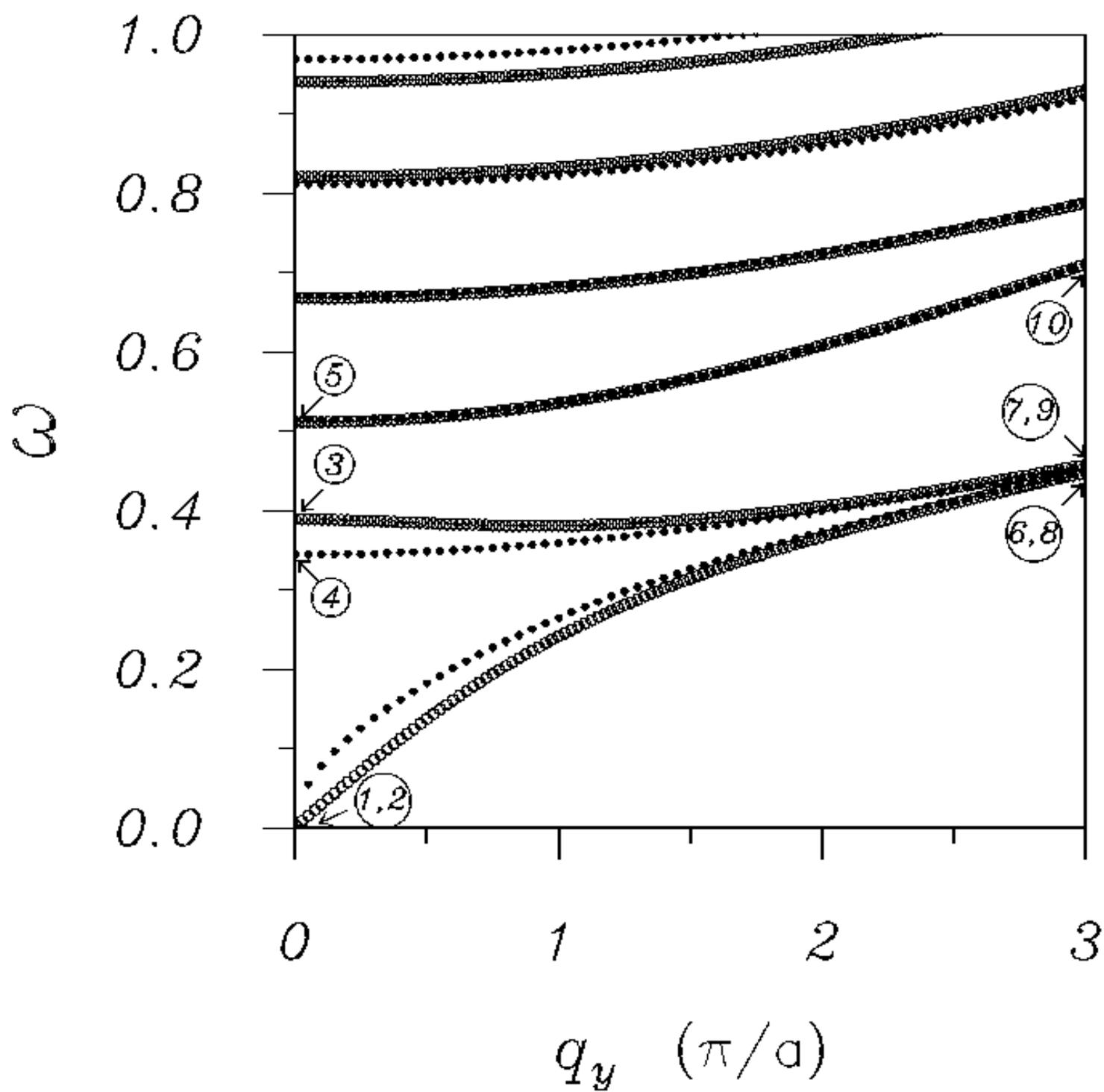
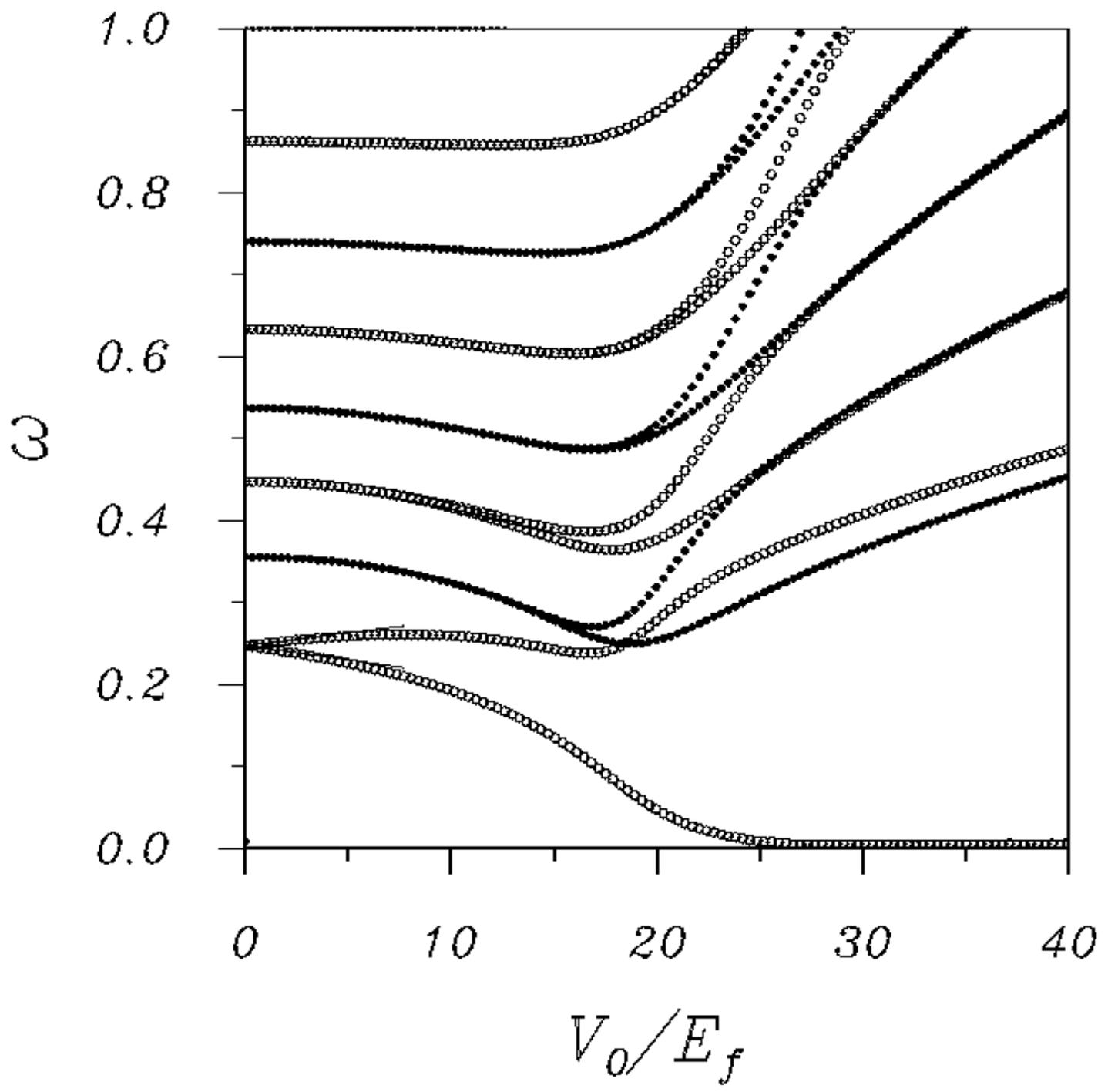
\end{document}